\documentclass[%
 reprint,
superscriptaddress,
 amsmath,amssymb,
 aps,
]{revtex4-2}

\usepackage{bbm}
\usepackage{graphicx}
\usepackage{dcolumn}
\usepackage{bm}
\usepackage{floatrow}
\usepackage{braket}
\usepackage{soul}
\graphicspath{ {./image/} }
\usepackage{hyperref}
      
\begin{document}

\preprint{APS/123-QED}
\title{Motional decoherence in ultracold Rydberg atom quantum simulators of spin models}

\author{ Zewen Zhang}
\affiliation{Department of Physics and Astronomy, Rice University, Houston, Texas 77005, USA}
\affiliation{ Rice Center for Quantum Materials, Rice University, Houston, Texas 77005, USA}
\author{ Ming Yuan}
\thanks{Current address: Pritzker School of Molecular Engineering, the University of Chicago, Chicago, Illinois 60637, USA}
\affiliation{Department of Physics and Astronomy, Rice University, Houston, Texas 77005, USA}
\affiliation{ Rice Center for Quantum Materials, Rice University, Houston, Texas 77005, USA}
\affiliation{School of the Gifted Young, University of Science and Technology of China, Hefei 230026, China}
\author{ Bhuvanesh  Sundar}
\thanks{Current address: Rigetti Computing}
\affiliation{Department of Physics and Astronomy, Rice University, Houston, Texas 77005, USA}
\affiliation{ Rice Center for Quantum Materials, Rice University, Houston, Texas 77005, USA}
\affiliation{JILA, Department of Physics, University of Colorado, Boulder, CO 80309, USA}
\author{ Kaden  R. A.  Hazzard}
\affiliation{Department of Physics and Astronomy, Rice University, Houston, Texas 77005, USA}
\affiliation{ Rice Center for Quantum Materials, Rice University, Houston, Texas 77005, USA}

\date{\today}

\begin{abstract}
Ultracold Rydberg atom arrays are an emerging platform for quantum simulation and computing. However, decoherence in these systems remains incompletely understood.  Recent experiments [Guardado-Sanchez \textit{et al.} Phys. Rev. X 8, 021069 (2018)] observed strong decoherence in the quench and longitudinal-field-sweep dynamics of two-dimensional Ising models  realized with ${}^6$Li Rydberg atoms in optical lattices. This decoherence was conjectured to arise from the atom motion couples to  electronic levels of atom. Here we show that the spin-motion coupling indeed leads to decoherence in qualitative, and often  quantitative,  agreement with the experimental data, treating the difficult spin-motion coupled problem using the discrete truncated Wigner approximation method. We also show that this decoherence will be an important factor to account for in future experiments with Rydberg atoms in optical lattices and microtrap arrays, and discuss methods to mitigate the effect of motion, such as using heavier atoms or deeper traps.
\end{abstract}

\maketitle

\section{\label{sec:level0}Introduction}

Arrays of  Rydberg atoms,   built either via optical lattices \cite{zeiher2016many,guardado2020quench,bharti2023picosecond} or programmable microtraps
\cite{labuhn2016tunable,de2019defect,gambetta2020engineering,bernien2017probing,kim2018detailed,lienhard2018observing,de2018accurate,scholl2020programmable,browaeys2020many,ebadi2021quantum,semeghini2021probing}, provide a versatile platform for studying quantum many-body systems, including
their real-time dynamics. Recent experiments have realized  synthetic quantum magnetic systems by constructing Rydberg-atom lattices with blockade radii $R_b$ that are comparable to the lattice constants $a_l$ \cite{de2018accurate,scholl2020programmable,browaeys2020many,song2020quantum,dauphin2016quantum,grusdt2018parton,ebadi2021quantum,bernien2017probing,kim2018detailed,semeghini2021probing,lienhard2018observing}. Such systems are widely used in areas like 
controlling many-body quantum phases \cite{browaeys2020many,dauphin2016quantum,song2020quantum,ebadi2021quantum,semeghini2021probing} and studying out-of-equilibrium dynamics \cite{bernien2017probing,lienhard2018observing,kim2018detailed,grusdt2018parton}.

While many of these experiments are conducted on  timescales where the atom motion is negligible~\cite{lienhard2018observing}, 
 recent experiments~\cite{guardado2018probing} have revealed significant discrepancies between the experimental results and the theoretical predictions even after accounting for  single-particle decoherence. 
 Although Ref.~\cite{guardado2018probing}   qualitatively explained this discrepancy by introducing a ``two-body interaction noise", their explanation does not predict the magnitude of the noise or its dependence on system parameters, and this magnitude  must be fit to the experimental data. Ref.~\cite{guardado2018probing} suggests the noise is likely to be an effect of atom motion, which
 becomes important on the timescale  $Jt\sim 1$, where $J$ is the strength of interaction between atoms.
 Although some theories of atom motion in Rydberg-atom systems have been developed, these have focused on few-atom motional effects or low-lying motional excitations~\cite{li2013nonadiabatic,macri2014rydberg,festa2021motion,robicheaux2021photon, magoni2022phonon}. Now, understanding motional effects in a many-body setting is crucial for an emerging generation of experiments~\cite{bernien2017probing, kim2018detailed, dudin2012strongly,baur2014single,muller2014implementation,han2016spectral,marcuzzi2017facilitation}. 

In this paper, we show that the experimentally observed deviations from 
the expected Ising model are consistent with the effects of  atom motion, which strongly supports atom motion as a source of Ref.~\cite{guardado2018probing}'s observed interaction noise, and our calculations allow us to predict its effects in a wide range of experiments.
There are two contributions from the atom motion: the first comes from the initial state's quantum and thermal spread of positions and momenta, and the second comes from the interaction-induced motion during the dynamics.  

We obtain these results by using the discrete truncated Wigner approximation (dTWA) \cite{schachenmayer2015many}, a semiclassical stochastic
phase space method, which circumvents the dificulties with direct numerical simulation that occur due to the large motional Hilbert space. 
As an aside, we note that Refs.~\cite{khasseh2020discrete,kunimi2021performance,czischek2018quenches} have applied dTWA to quenches, while our results include both quench and ramp dynamics.

The prediction of motional decoherence in Rydberg atom arrays
is especially important for the future of the Rydberg atom array quantum simulation platform. We show that  while working with optical tweezers instead of a lattice reduces the motional decoherence, the decoherence nevertheless
can be significant on timescales currently being experimentally explored~\cite{bernien2017probing,scholl2020programmable,guardado2020quench}.  Also, our calculations quantify how the decoherence depends on the depth of the tweezers or optical lattice, the atomic mass, and the sign of the interactions, providing a road map to mitigating the decoherence. 

Our paper is structured as follows. Sec. \ref{sec:model} introduces the spin-motion coupled Rydberg-atom lattice model and the dTWA method. Sec. \ref{sec:results} shows the dynamics following  sudden quenches and slow detuning ramps, comparing dTWA  with and without motion.
Sec. \ref{sec:extension} quantifies how the decoherence depends on  experimental settings in optical lattices and microtraps. Sec. \ref{sec:con} concludes.

\section{\label{sec:model}Model and method}

This Section presents the spin-motion coupled Ising model used to describe experiments of Rydberg atoms in optical lattices or microtrap arrays [Sec.~\ref{sec:smc}] and the 
dTWA methods used to calculate the  dynamics [Sec.~\ref{sec:twa}].

\subsection{\label{sec:smc}Spin-motion Coupled Ising Model}

\begin{figure}[h]
    \centering
    \includegraphics[width=0.8\columnwidth]{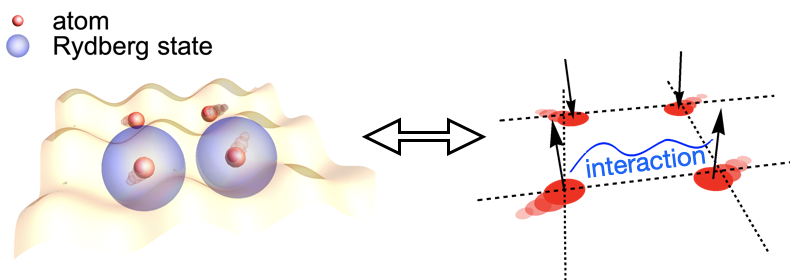}
    \caption{\label{fig:model}(color online) 
    A 2D optical lattice of atoms with two internal states coupled by a laser realizes a transverse-field Ising model with $\ket{\downarrow}$ and $\ket{\uparrow}$  given by the ground state $\ket{g}$ and a Rydberg state $\ket{r}$, respectively. Two  atoms  interact only if both are in $\ket{\uparrow}$, as described by Eq. \eqref{hamiltonian}.
   Spins (atoms) are allowed to move, as indicated by the pale red trails.
    }
\end{figure}

The experimental system we consider is an array of atoms  in a 2D square lattice at unit filling driven by  a laser that  couples ground state atoms $\ket{g}=\ket{\downarrow}$ to a low-lying Rydberg state $\ket{r}=\ket{\uparrow}$, with Rabi frequency $\Omega$  and detuning $\Delta$, as shown in Fig.~\ref{fig:model}. 
In the spin language, the $\Omega$ and $\Delta$ give transverse and longitudinal fields, respectively. 
There is an attractive  van der Waals (vdW) interaction when two atoms are in the Rydberg state, giving rise to an Ising coupling between spins that is $C_6/R^6$, where $R$ is the distance between spins. The nearest-neighbor terms dominate for the parameters in Ref.~\cite{guardado2018probing}, so we neglect longer-ranged  interactions.

Atoms  move due to their initial velocities, the vdW interaction among those in the Rydberg state, and the trapping potential. Therefore, the Ising model must be extended to capture 
the kinetic energy and the optical forces on the atoms. We assume  atom $i$ is trapped by  a 3D harmonic potential centered at lattice site $i$,  which is valid in a deep lattice  as long as the atoms are not too far displaced from their equilibrium position $\vec{R}_i$.  Then the appropriate Hamiltonian is  
\begin{align}
 H = &  \frac{\hbar}{2}\sum_i\left(\Omega \sigma_i^x -\Delta \sigma_i^z \right) \nonumber \\
 &{}-   \frac{\hbar J}{4}\sum_{\langle i, j\rangle}\left(\frac{a_l}{r_{ij}}\right)^6(\sigma_i^z+1)(\sigma_j^z+1)\nonumber\\
   &{}+\sum_{i,\alpha}\frac{{p_i^\alpha}^2}{2m}+  \sum_i\hbar V_0\frac{\pi^2}{2a_l^2}\tilde{r_i}^2(1-\sigma_i^z),
\label{hamiltonian}
\end{align}
where $a_l$ is the  lattice constant, $J=|C_6|/a_l^6$ is the nearest-neighbor interaction strength, $p_i^\alpha$ is atom $i'$s momentum  in direction $\alpha$  ($\alpha=x,y,z$), $\sigma_i^\alpha$ is the Pauli matrix for atom $i$, $r_{ij}$ is the distance between two atoms $i$ and $j$ that are located at $\vec{r}_i$ and $\vec{r}_j$, and  $\tilde{r}_i$ is the magnitude of the displacement $\vec{\tilde{r}}_i=\vec{r}_i-\vec{R}_i$ of  atom $i$ from  the center of   lattice site.  As indicated by the last term, the trapping potential is treated only on ground-state atoms and neglected on Rydberg-state atoms, as on Rydberg-state atom the trap is much smaller than the trap on ground-state atoms, as well as negligible compared to the Rydberg-Rydberg interaction.
 
Most of our concrete calculations are for Lithium-6 atoms in an optical lattice in the experimental configuration of Ref.~\cite{guardado2018probing}, but the broad results apply more generally, for example to other atoms and microtrap potentials, which are discussed in  Sec.~\ref{sec:extension}.  
When considering  the experiments in Ref.~\cite{guardado2018probing},
we take the lattice depth to be $V_0=55E_R/\hbar$, where $E_R=\pi^2 \hbar^2/2ma_l^2$ is the recoil energy and  $a_l=1024/\sqrt{2}$nm, and  $C_6/a_l^6 =-2\pi\times6.0$MHz, the result from experimental fitting in Ref.~\cite{guardado2018probing}.
Throughout our calculation,   we neglect the single-particle decoherence  
unless otherwise specified. To measure how the system deviates from product states, we focus on the connected spin-spin correlation 
\begin{equation}
    C(i,j)=\langle \sigma_i^z\sigma_j^z\rangle - \langle \sigma_i^z\rangle\langle \sigma_j^z\rangle.
\end{equation}

\subsection{\label{sec:twa}Discrete Truncated Wigner Approximation}

In the area of quantum spin dynamics, 
the  truncated Wigner approximation 
(TWA) methods are widely used, especially for quenches and when timescales are not too long.
These methods are controlled approximations when the correlations between particles are small or the system is semiclassical~\cite{polkovnikov2010phase}.
Additionally, the dTWA method\cite{schachenmayer2015many,pucci2016simulation,acevedo2017exploring,sundar2019analysis},
which maps the spin degree of freedom  to a discrete phase space, can also preserve oscillating features~\cite{czischek2018quenches,kunimi2021performance} and capture correlations~\cite{khasseh2020discrete} on moderate timescales; often, it  outperforms the continuous TWA in 1D spin chains~\cite{sundar2019analysis}. Recent  studies \cite{kunimi2021performance,schachenmayer2015dynamics,signoles2021glassy,schuckert2020nonlocal} compare results from dTWA to exact results, higher order dTWA calculations, or Rydberg atom experiments and reveal that dTWA methods in 2D spin systems can be accurate for timescales of $Jt\simeq 2$.  
TWA methods approximate quantum dynamics by  classical equations of motion (EoM) propagated from the exact initial state's Wigner  distribution (quasiprobability)~\cite{polkovnikov2010phase}. The EoM's are derived from the Hamiltonian~\eqref{hamiltonian} and shown in Appendix~\ref{appendix:eom}. We use the same random seed for calculations with different physical parameters.

For atoms in a harmonic trap with frequency $\omega$ at temperature $T$, the Wigner distribution is
\begin{equation}
\label{eq:thermal_initial}
W(x,p)=
 2\tanh\left(\frac{\hbar\omega}{2k_BT}\right) \exp\left(-\frac{(x/\ell)^2+(p\ell/\hbar)^2}{\coth(\frac{\hbar \omega}{2k_B T})}\right),
\end{equation}
where $\ell=\sqrt{\hbar/m\omega}$ is the  harmonic oscillator length.
$T=0$ yields the ground-state Wigner distribution.
For the spins, 
dTWA represents the spin-$\frac{1}{2}$ observables as points in a discrete phase space \cite{wootters1987wigner,schachenmayer2015many},  
 which we take to be the eight points $\vec{S}_u \in \frac{1}{2}(\pm 1,\pm 1,\pm 1)$. Ref.~\cite{sundar2019analysis} has shown that this choice of basis performs better than other choices in a broad range of scenarios~\cite{pucci2016simulation}.
When the initial spins are aligned in the $\ket{\downarrow}$ state, the quasiprobability $W(\vec{S}_u)$ is $\frac{1}{4}$ for the four points $\frac{1}{2}(\pm 1,\pm 1,- 1)$ and $0$ for the  other four.

Our dTWA and exact diagonalization (ED) calculations are for a $4\times 4$ lattice with periodic boundary condition. We have calculated results for a  $6\times 6$ system with dTWA and they  agree well with the $4\times 4$ results. 
 For example, for the long-time quench results  discussed in Sec.~\ref{sec:long}, the differences of the correlation between two system sizes are always less  (usually much less) than  6\% for the range of  $\Delta$'s plotted, and   for shorter times these differences are  smaller. 
For both ED dynamics and solving the classical EoM,  we use  fourth-order Runge–Kutta  with a timestep of 1ns for the ideal Ising model and 0.5ns for the motional model. The results are well-converged: for example, decreasing the integration  timestep by a factor of two for the ideal Ising long-time quench discussed in Sec.~\ref{sec:long} changes all results by less than 2\%. Therefore, in view of the comparison to ED methods and larger system sizes, the dTWA method we applied here is reliable to simulate the spin dynamics combined with motional degrees of freedom. Statistical errors are small and indicated in the figures.

\section{Results: Quench and ramp dynamics}
\label{sec:results}
\subsection{Sudden Quench}
\label{sec:sudden}

Motivated by Ref.~\cite{guardado2018probing}, the first scenario we study is  quench dynamics of the Rydberg atoms in an optical lattice. Specifically, the lattice is initially   filled with exactly one atom per site, which is in its electronic ground state (spin-down) and in its motional ground state. (Sec.~\ref{sec:extension} considers finite-temperature motional states as well.) After an instantaneous  turn-on of $\Omega$, the system evolves under Eq.~\eqref{hamiltonian}. We study two quenches: ``short-time quenches"   with $\Omega t < \pi$ (Sec.~\ref{sec:short}) and ``long-time quenches" with $\Omega t >\pi$ (Sec.~\ref{sec:long}). For each quench, we calculate the spin correlations $C(i,j)$ as a function of detuning $\Delta$.

\subsubsection{Short-time Quench\label{sec:short}}

\begin{figure}[h]
\includegraphics[width=0.92\linewidth]{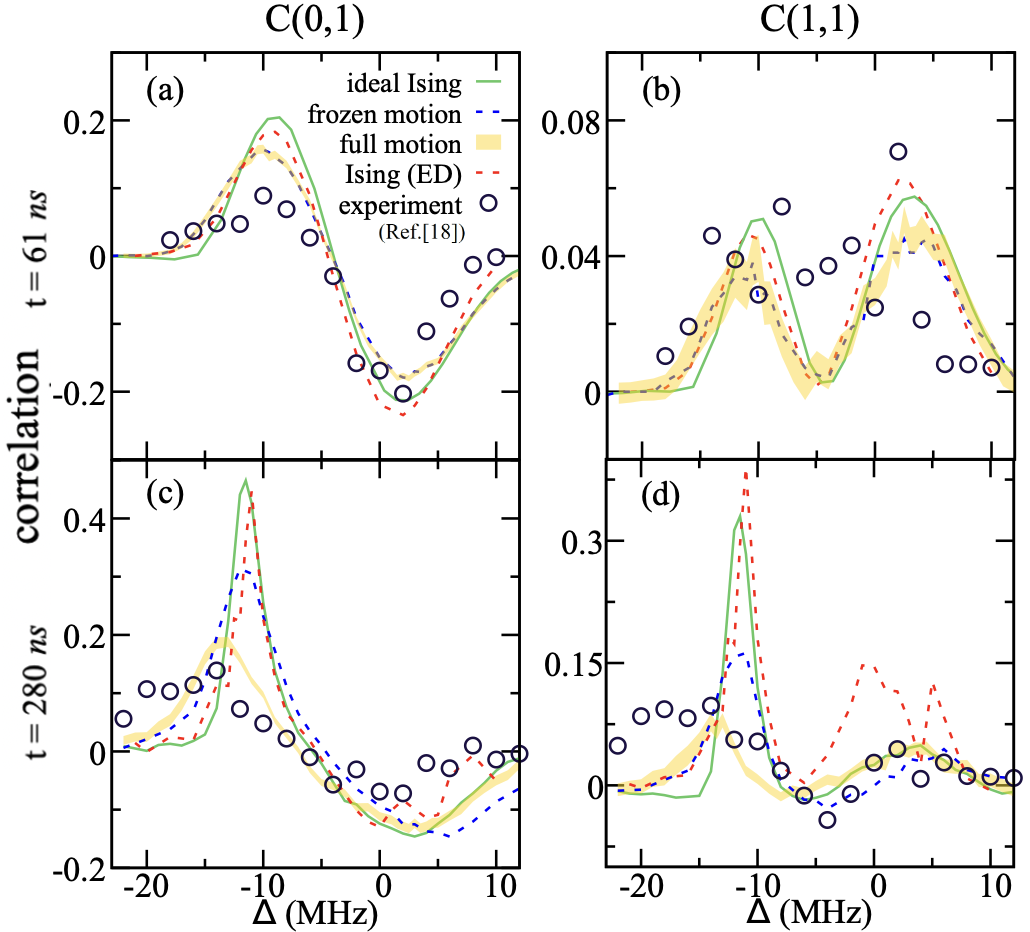}
\caption{(color online)  Spin correlations at short duration (top row, $\Omega t=0.5\pi$, 61ns,  5,000 dTWA trajectories) and at long duration (bottom row, $\Omega t=2.97\pi$, 280ns,  2,000 dTWA trajectories) after a sudden quench with different  $\Delta$ values. The circles are experimental data from  Fig.~2 in Ref.~\cite{guardado2018probing} and the curves are theory calculations described in the   text. 
The shaded region  indicates $\pm 1$ standard error of the mean, and to avoid clutter are shown only on the full motion results, which are the results with the  largest statistical errors. In (b), the disagreement between the simulation and the experiment is mainly caused by the small magnitude of correlation. }
\label{fig:quench}
\end{figure}

Figures~\ref{fig:quench}(a) and (b) show that the effect of motion is small but observable for the short-time quench ($\Omega t=0.5\pi$) dynamics. 
In these calculations, we use the   Rabi frequency ($\Omega=2\pi\times 4.1$MHz) and duration ($t=61$ns) from the experiments in Ref.~\cite{guardado2018probing}.  We compare three versions of the spin model: (\textit{i}) the ideal Ising model with all spins sitting at the center of individual lattice sites; (\textit{ii}) the frozen model, where each atom has a finite-width distribution of locations initially, according to  Eq.~\eqref{eq:thermal_initial}, but does not move during the spin dynamics; (\textit{iii}) the full motion  model, where each atom has both an initial distribution of locations and momenta, and it is allowed to move according to interactions with other atoms and the optical potential.

Figures~\ref{fig:quench}(a) and (b) shows that the motion modifies the correlations  even on these short timescales, as seen by comparing the ideal Ising and the full motion model.
However, for this timescale, the dominant contribution comes from initial spread of atom positions rather than atom motion during the quench,    evidenced by   the close agreement between the full-motion and frozen-motion  results.

Figure~\ref{fig:quench} also shows results obtained by $4\times 4$ ED of the pure spin system. Comparing this and the dTWA method confirms that the latter is an accurate approximation in the  situations shown.

\subsubsection{Long-time Quench}
\label{sec:long}

Figures~\ref{fig:quench}(c) and (d)  show that atom motion significantly suppresses the growth of correlations in the long-time quench dynamics with $\Omega t = 2.97\pi$, using the corresponding experimental 
$\Omega=2\pi\times 5.4$MHz and $t=280$ns.

 The simulations with atom motion qualitatively reproduce the experimental long-time quench results. 
Upon incorporating the initial state effects and motion during the dynamics (the full motion model), the height of the correlation peaks is dramatically reduced, in agreement  with the experimental values. The atom motion also broadens and shifts the peaks. 
We also find that the frozen motion model deviates strongly from  the full motion results. While some wiggles shown in the ED results do not appear in the dTWA results, our conclusion about the motional effects stands when comparing the motional results to coherent results.

Comparing short and long duration quenches demonstrates that the effects of motion increase over time.
The evolution of correlations in both ideal Ising and full motion models, as shown in Appendix~\ref{appendix:time},  confirms that the effect of motion accumulates over time and shows up after $t=50$ns in this experimental setting, independent of the detuning. A drastic increasing of kinetic energy, specifically in the transverse direction, is also observed in Appendix~\ref{appendix:time}.

\subsection{Slow Quench with Detuning Ramp}
\label{sec:ramping}
Again motivated by Ref.~\cite{guardado2018probing}, the second scenario we study is the dynamics as the  longitudinal field $\Delta$ is slowly swept from far off resonance to the final detuning. For sufficiently slow ramps, coherent evolution, and no motional coupling, this would result in adiabatic preparation of the Ising ground state.  The dynamics has two stages, as the first stage in the three-stage dynamics in Fig. 3 of Ref. \cite{guardado2018probing} is omitted for simplicity. This simplification has little to do with our estimate of the motional effect, as the large detuning in this stage prevents atoms from excited to the Rydberg state.
Initially, all the atoms are prepared in the ground state under a laser field with Rabi frequency $\Omega_i$ and large laser detuning $\Delta_i\gg J$. The first stage of the ramping process is that the $\Delta$ decreases to a preset value $\Delta_f$ with a given ramp rate $\dot{\Delta}$ while  $\Omega$ is constant. The second stage is a linear turn-off of the $\Omega$ with rate $\dot{\Omega}$ while $\Delta$ holds constant. We focus on the experimental parameter values:  $J=2\pi\times 6$MHz, $\Omega_i=0.9J$, $\Delta_i = 3.3J$, and
$\dot{\Omega}=0.24 J^2$, and we take the single-body decoherence times, specifically the lifetime and dephasing time, to be  $T_1=20\mu$s and $T_2=0.5\mu$s, respectively. The single-body decoherence is incorporated into the calculations as described in Appendix~\ref{appendix:eom}.

\begin{figure}[h]
    \centering
    \includegraphics[width=0.8\columnwidth]{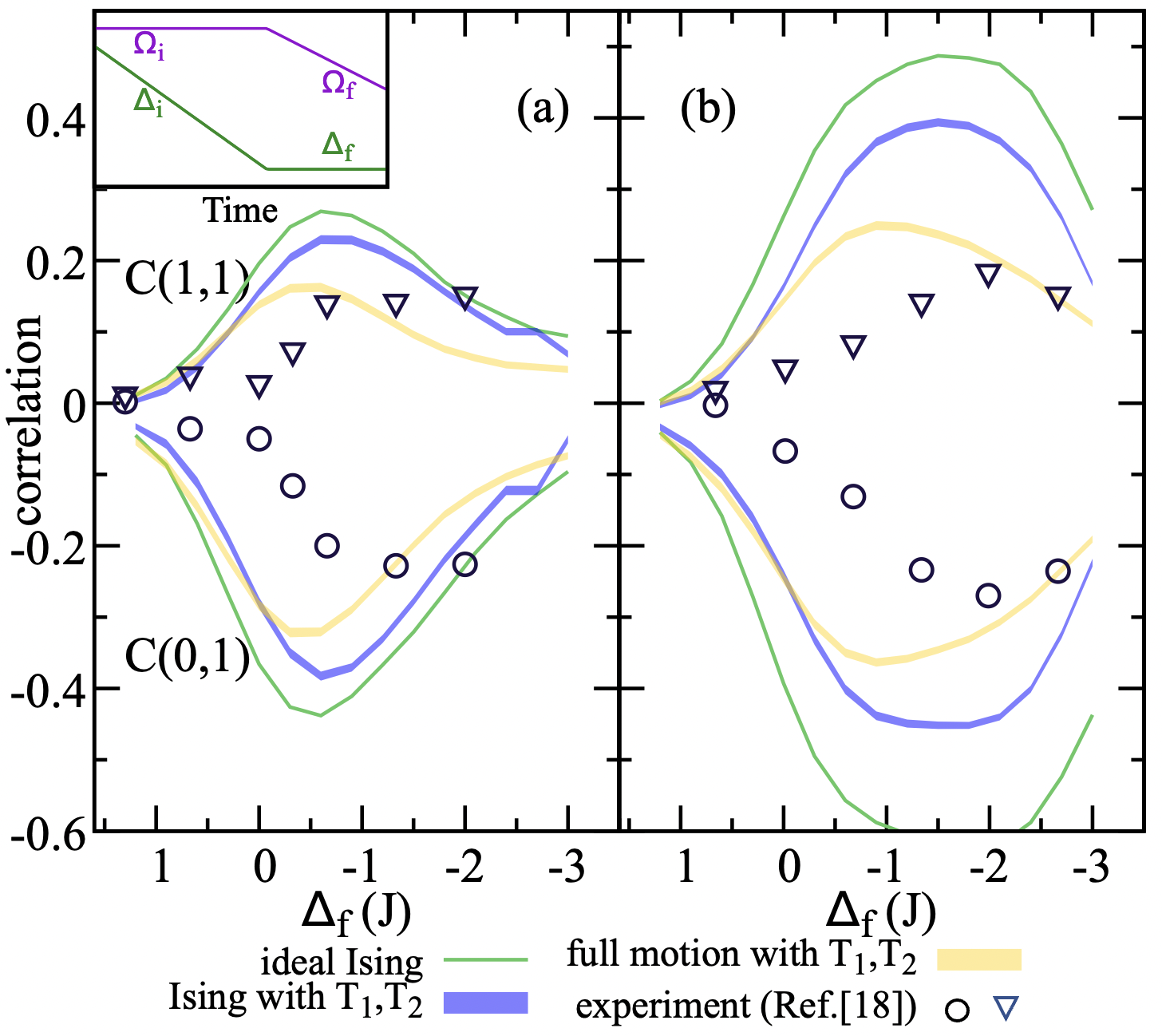}
     \caption{(color online)  Ramp results varying $\Delta_f$.  (a)~$\dot{\Delta}=-1.42 J^2$. The  fields $\Omega$ and $\Delta$  vary over time in two stages, as  shown in the inset and described in the text.(b)~$\dot{\Delta}=-0.70J^2$. 
    The  green curves are the ideal Ising results. The two shaded regions, which indicate $\pm 1$ standard error of the mean, are Ising and full motion model results (blue and yellow curves, respectively), each with single-body $T_1$ and $T_2$ decoherence.  Note that the horizontal axis is reversed. The circles and triangles are experimental data from  Fig.~3 in Ref. \cite{guardado2018probing}
     }
    \label{fig:ramp}
\end{figure}

Figure~\ref{fig:ramp} shows  results for two ramp  rates $\dot{\Delta}$, in order  to compare  motional effects at different timescales. For  the shortest duration ramps [Fig.~\ref{fig:ramp}(a) at  positive $\Delta_f$], the single-atom noise  can play a more important role than the atom motion. However, the effect of motion rapidly appears and exceeds the single-atom noise, as seen in Fig.~\ref{fig:ramp}(a) for negative $\Delta_f$ and Fig.~\ref{fig:ramp}(b) for nearly all $\Delta_f$.  
In Fig.~\ref{fig:ramp}(b), as the sweeping is slower (with smaller $\dot{\Delta}$), the process will be closer to an adiabatic process and build up a higher correlation peak. However, as this process takes longer time, the effect of the atom motion is even larger.

The dTWA simulation supports  Ref.~\cite{guardado2018probing}'s conjecture that  atom motion is a significant source of decoherence in the detuning ramp results. Even though there are some differences between simulation and experiments that are possibly caused either by the limit of the dTWA method or the simplification in the model, the numerical results agree roughly with the  experimentally-observed correlations as a function of detuning \cite{guardado2018probing}, and give a peak magnitude quantitatively consistent with the experiment. As the atom motion is introduced to the model without any fitting, our results give a reasonable explanation of Ref.~\cite{guardado2018probing} two-body interaction noise.

\section{Effect of Motion in Optical Tweezers}
\label{sec:extension}

 The results above have shown the importance of motion in recent optical lattice experiments, 
 raising the question of what role  atom motion plays more generally in ongoing experiments with optical lattices and microtraps, another important technique to realize spin models with Rydberg atoms~\cite{scholl2020programmable,ebadi2021quantum}. In this section, we  answer two relevant questions: First, what are possible ways to mitigate the motional effects in optical lattices? Second, what is the effect of atom motion in  optical tweezer experiments? 

\subsection{Suppressing  motional decoherence }
\begin{figure}
    \includegraphics[width=0.45\columnwidth]{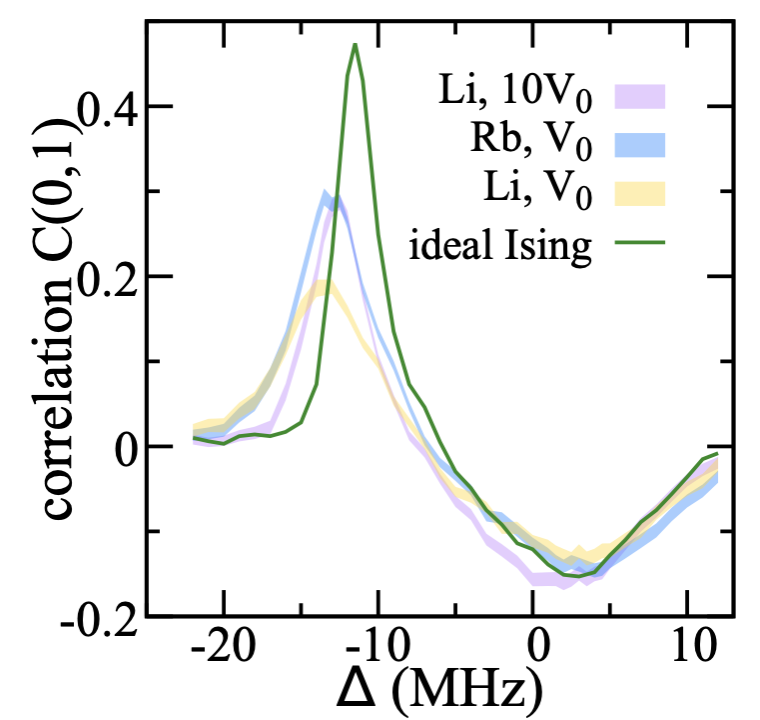}
    \caption{\label{fig:reduce} (color online) Long time ($\Omega t = 2.97\pi$) quench when Rb atoms are used or the depth of the lattice increases to $10V_0$, where $V_0$ is the lattice depth applied in Sec. \ref{sec:results}. These two results are plotted in comparison to the ideal Ising and full motion  results of the original setting. The shaded region  indicates $\pm 1$ standard error of the mean. 
    }
\end{figure}

Intuitively, the effect of motion will be smaller in deeper lattice or with heavier atoms. Our calculations show that  replacing lithium atoms with rubidium atoms or increasing the lattice depth can help to reduce, but not eliminate, the motional effects, as shown in Fig. \ref{fig:reduce}.
In  dynamics similar to the long-time quench in Sec.~\ref{sec:long} ($\Omega t =2.97\pi$),  using heavier atoms (Rb) or applying a deeper trap ($V=10V_0$) increases the height of correlation peaks and is closer to the ideal Ising result. However, the peak remains much lower than in the ideal Ising dynamics and shifts noticeably. Because these techniques  only partially alleviate the motional decoherence, atom motion will be important to consider when designing  future experiments.

\subsection{Motional decoherence in microtraps}
Although the programmability of microtraps leads to a diversity of geometries, the effect of motion is generally expected to be smaller in microtraps than lattices, due to the microtraps' deeper wells. 
To evaluate these effects, we simulate the sudden quench dynamics with atom motion in microtraps in  recently-used experimental conditions with $^{87}$Rb \cite{scholl2020programmable,barredo2018synthetic}. The parameters include $a_l=10\mu$m, $C_6=\pm 2\pi \times 1.95$MHz$a_l^6$,  $\Omega=2\pi \times 1.95$MHz, and a radial (longitudinal) harmonic oscillator frequency near the trap center of 100kHz (20kHz).

\begin{figure}[h]
\includegraphics[width=0.75\linewidth]{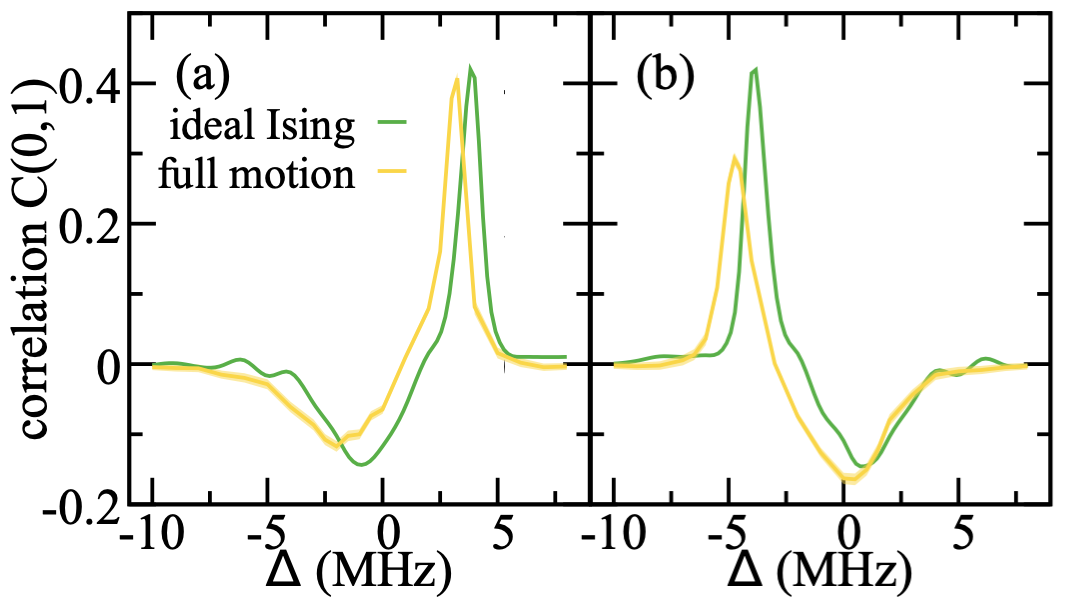}
\caption{ (color  online) 
Sudden quench results for Rydberg-atom lattice based on Rb atoms trapped in microtraps, under different detuning values. The quench time is 760ns ($\Omega t=2.97\pi$). 
(a) C(0,1) with repulsive vdW interaction; (b) C(0,1)  with attractive interaction.
The shaded region  indicates $\pm 1$ standard error of the mean.
 \label{fig:microtrap} }
\end{figure}

 The quench duration is 760ns, which gives the same $\Omega t$ value as the long-time quench discussed in Sec.~\ref{sec:sudden}.
The results are shown in Fig. \ref{fig:microtrap}, along with the dTWA results of the ideal Ising model for comparison. It should be noted that the atoms in microtraps are not fully cooled down after the preparation of the experiment. Thus, the initial conditions [Eq.~\eqref{eq:thermal_initial}]
correspond to the atoms' temperature in the magneto-optical trap, which is 25$\mu$K
\cite{barredo2018synthetic}.

Figure~\ref{fig:microtrap}(a) shows that in microtraps with repulsive dipole-dipole interactions, a common case in  experiments~\cite{scholl2020programmable,ebadi2021quantum}, the motional effects are fairly small in the timescale we considered.
After the sudden quench, the full motion model's peak in $C(0,1)$ 
has the same height as in the ideal Ising model, but with a slightly shifted position. 

An important point is that  in ongoing  microtrap experiments,  $Jt$ is  10 times larger than we consider here~\cite{scholl2020programmable}. We have performed dTWA calculations for such timescales, seeing a much larger motional effect. This is not included in the figures because the strong quantum fluctuations and correlations at the longer times  make our dTWA method questionable.  Nevertheless, our results here suggest that the effect of motional decoherence should already be  detectable at much shorter times, and, as discussed in Sec.~\ref{sec:long}, it is reasonable to expect an increase in  this motional effect at longer times.

Finally,  we note that the sign of the interaction 
  changes the motional effects as shown in Fig. \ref{fig:microtrap}(b). The effect of motion for repulsive interactions  is less significant than that for  attractive interactions.
The  $C(0,1)$ in the attractive Rydberg-atom lattice [Fig. \ref{fig:microtrap}(b)] shows a suppression of peak height by the atom motion, which is not observed in the repulsive model in Fig. \ref{fig:microtrap}(a). The atom motion in the attractive model also shifts the peak position.

\section{Conclusion}
\label{sec:con}

In summary, we show that atom motion affects  
the evolution of correlations  in Rydberg-atom lattices, 
which   explains 
the discrepancy between coherent results and recent experimental observations. 
For example, the results indicate that  atom motion suppresses magnetic correlations,  sometimes by a large amount.

We also suggest possible ways to relieve the effect of atom motion including using heavier atoms, deeper lattices, or repulsive Rydberg-Rydberg interactions. All of these methods, which are usable in current experiments, can  reduce, but not eliminate, the motional decoherence on the timescales we have considered.
As the Rydberg atom platforms become a major toolbox in quantum simulation and quantum computation, knowing the existence of motional decoherence, especially its magnitude in different timescales, is helpful both to design experiments and analyze results. Another route worth investigating to suppress motional decoherence is to use ultrafast dynamics, as in recent experiments in Refs.~\cite{takei2016direct,sommer2016time,bharti2023picosecond}. Here, the short timescales minimize the thermal motion. If interaction-induced motion also remains slow compared to the ultrafast timescales, this platform may offer a route to simulate long-time spin dynamics with minimal motional effects.

Interesting future questions include  how the motional dynamics interplays with  other sources of  noise including disorder in the lattice~\cite{scholl2020programmable}, blackbody-induced broadening \cite{goldschmidt2016anomalous,boulier2017spontaneous} or transitions \cite{aman2016trap}, motion-facilitated excitation \cite{festa2021motion,boulier2017spontaneous} or interaction more than two-body. We have noticed that some quantitative differences remain between the dTWA simulation and experiments. To understand this, one may look to improve the motional Rydberg-atom lattice model to include effects of longer-range interaction \cite{guardado2018probing}, shapes of Rydberg-Rydberg interaction \cite{macri2014rydberg}, and to extend the numerical methods to capture the strong quantum fluctuations that occur at 
timescales \cite{czischek2018quenches,wurtz2018cluster} that beyond the current limit of dTWA methods.

\begin{acknowledgments} 
We thank Fei Gao for conversations about related calculations.
We acknowledge support by the
Welch Foundation  Grant No. C1872, the Office of Naval Research Grant No. N00014-20-1-2695,
and the National Science Foundation  Grant No.
PHY1848304. K.H.'s contribution benefited from discussions at the KITP, which was supported in part by the National Science Foundation under Grant No. NSF PHY-1748958.

\end{acknowledgments}

\bibliography{apssamp}

\providecommand{\noopsort}[1]{}\providecommand{\singleletter}[1]{#1}%
\begin{thebibliography}{55}%
\makeatletter
\providecommand \@ifxundefined [1]{%
 \@ifx{#1\undefined}
}%
\providecommand \@ifnum [1]{%
 \ifnum #1\expandafter \@firstoftwo
 \else \expandafter \@secondoftwo
 \fi
}%
\providecommand \@ifx [1]{%
 \ifx #1\expandafter \@firstoftwo
 \else \expandafter \@secondoftwo
 \fi
}%
\providecommand \natexlab [1]{#1}%
\providecommand \enquote  [1]{``#1''}%
\providecommand \bibnamefont  [1]{#1}%
\providecommand \bibfnamefont [1]{#1}%
\providecommand \citenamefont [1]{#1}%
\providecommand \href@noop [0]{\@secondoftwo}%
\providecommand \href [0]{\begingroup \@sanitize@url \@href}%
\providecommand \@href[1]{\@@startlink{#1}\@@href}%
\providecommand \@@href[1]{\endgroup#1\@@endlink}%
\providecommand \@sanitize@url [0]{\catcode `\\12\catcode `\$12\catcode `\&12\catcode `\#12\catcode `\^12\catcode `\_12\catcode `\%12\relax}%
\providecommand \@@startlink[1]{}%
\providecommand \@@endlink[0]{}%
\providecommand \url  [0]{\begingroup\@sanitize@url \@url }%
\providecommand \@url [1]{\endgroup\@href {#1}{\urlprefix }}%
\providecommand \urlprefix  [0]{URL }%
\providecommand \Eprint [0]{\href }%
\providecommand \doibase [0]{https://doi.org/}%
\providecommand \selectlanguage [0]{\@gobble}%
\providecommand \bibinfo  [0]{\@secondoftwo}%
\providecommand \bibfield  [0]{\@secondoftwo}%
\providecommand \translation [1]{[#1]}%
\providecommand \BibitemOpen [0]{}%
\providecommand \bibitemStop [0]{}%
\providecommand \bibitemNoStop [0]{.\EOS\space}%
\providecommand \EOS [0]{\spacefactor3000\relax}%
\providecommand \BibitemShut  [1]{\csname bibitem#1\endcsname}%
\let\auto@bib@innerbib\@empty
\bibitem [{\citenamefont {Zeiher}\ \emph {et~al.}(2016)\citenamefont {Zeiher}, \citenamefont {Van~Bijnen}, \citenamefont {Schau{\ss}}, \citenamefont {Hild}, \citenamefont {Choi}, \citenamefont {Pohl}, \citenamefont {Bloch},\ and\ \citenamefont {Gross}}]{zeiher2016many}%
  \BibitemOpen
  \bibfield  {author} {\bibinfo {author} {\bibfnamefont {J.}~\bibnamefont {Zeiher}}, \bibinfo {author} {\bibfnamefont {R.}~\bibnamefont {Van~Bijnen}}, \bibinfo {author} {\bibfnamefont {P.}~\bibnamefont {Schau{\ss}}}, \bibinfo {author} {\bibfnamefont {S.}~\bibnamefont {Hild}}, \bibinfo {author} {\bibfnamefont {J.-y.}\ \bibnamefont {Choi}}, \bibinfo {author} {\bibfnamefont {T.}~\bibnamefont {Pohl}}, \bibinfo {author} {\bibfnamefont {I.}~\bibnamefont {Bloch}},\ and\ \bibinfo {author} {\bibfnamefont {C.}~\bibnamefont {Gross}},\ }\bibfield  {title} {\bibinfo {title} {Many-body interferometry of a {R}ydberg-dressed spin lattice},\ }\href@noop {} {\bibfield  {journal} {\bibinfo  {journal} {Nat. {P}hys.}\ }\textbf {\bibinfo {volume} {12}},\ \bibinfo {pages} {1095} (\bibinfo {year} {2016})}\BibitemShut {NoStop}%
\bibitem [{\citenamefont {Guardado-Sanchez}\ \emph {et~al.}(2020)\citenamefont {Guardado-Sanchez}, \citenamefont {Spar}, \citenamefont {Schauss}, \citenamefont {Belyansky}, \citenamefont {Young}, \citenamefont {Bienias}, \citenamefont {Gorshkov}, \citenamefont {Iadecola},\ and\ \citenamefont {Bakr}}]{guardado2020quench}%
  \BibitemOpen
  \bibfield  {author} {\bibinfo {author} {\bibfnamefont {E.}~\bibnamefont {Guardado-Sanchez}}, \bibinfo {author} {\bibfnamefont {B.}~\bibnamefont {Spar}}, \bibinfo {author} {\bibfnamefont {P.}~\bibnamefont {Schauss}}, \bibinfo {author} {\bibfnamefont {R.}~\bibnamefont {Belyansky}}, \bibinfo {author} {\bibfnamefont {J.~T.}\ \bibnamefont {Young}}, \bibinfo {author} {\bibfnamefont {P.}~\bibnamefont {Bienias}}, \bibinfo {author} {\bibfnamefont {A.~V.}\ \bibnamefont {Gorshkov}}, \bibinfo {author} {\bibfnamefont {T.}~\bibnamefont {Iadecola}},\ and\ \bibinfo {author} {\bibfnamefont {W.~S.}\ \bibnamefont {Bakr}},\ }\bibfield  {title} {\bibinfo {title} {Quench dynamics of a {F}ermi gas with strong long-range interactions},\ }\href@noop {} {\bibfield  {journal} {\bibinfo  {journal} {Energy}\ }\textbf {\bibinfo {volume} {10}},\ \bibinfo {pages} {1} (\bibinfo {year} {2020})}\BibitemShut {NoStop}%
\bibitem [{\citenamefont {Bharti}\ \emph {et~al.}(2023)\citenamefont {Bharti}, \citenamefont {Sugawa}, \citenamefont {Mizoguchi}, \citenamefont {Kunimi}, \citenamefont {Zhang}, \citenamefont {de~L{\'e}s{\'e}leuc}, \citenamefont {Tomita}, \citenamefont {Franz}, \citenamefont {Weidem{\"u}ller},\ and\ \citenamefont {Ohmori}}]{bharti2023picosecond}%
  \BibitemOpen
  \bibfield  {author} {\bibinfo {author} {\bibfnamefont {V.}~\bibnamefont {Bharti}}, \bibinfo {author} {\bibfnamefont {S.}~\bibnamefont {Sugawa}}, \bibinfo {author} {\bibfnamefont {M.}~\bibnamefont {Mizoguchi}}, \bibinfo {author} {\bibfnamefont {M.}~\bibnamefont {Kunimi}}, \bibinfo {author} {\bibfnamefont {Y.}~\bibnamefont {Zhang}}, \bibinfo {author} {\bibfnamefont {S.}~\bibnamefont {de~L{\'e}s{\'e}leuc}}, \bibinfo {author} {\bibfnamefont {T.}~\bibnamefont {Tomita}}, \bibinfo {author} {\bibfnamefont {T.}~\bibnamefont {Franz}}, \bibinfo {author} {\bibfnamefont {M.}~\bibnamefont {Weidem{\"u}ller}},\ and\ \bibinfo {author} {\bibfnamefont {K.}~\bibnamefont {Ohmori}},\ }\bibfield  {title} {\bibinfo {title} {Picosecond-scale ultrafast many-body dynamics in an ultracold {R}ydberg-excited atomic {M}ott insulator},\ }\href@noop {} {\bibfield  {journal} {\bibinfo  {journal} {Phys. {R}ev. {L}ett.}\ }\textbf {\bibinfo {volume} {131}},\ \bibinfo {pages} {123201} (\bibinfo {year} {2023})}\BibitemShut {NoStop}%
\bibitem [{\citenamefont {Labuhn}\ \emph {et~al.}(2016)\citenamefont {Labuhn}, \citenamefont {Barredo}, \citenamefont {Ravets}, \citenamefont {De~L{\'e}s{\'e}leuc}, \citenamefont {Macr{\`\i}}, \citenamefont {Lahaye},\ and\ \citenamefont {Browaeys}}]{labuhn2016tunable}%
  \BibitemOpen
  \bibfield  {author} {\bibinfo {author} {\bibfnamefont {H.}~\bibnamefont {Labuhn}}, \bibinfo {author} {\bibfnamefont {D.}~\bibnamefont {Barredo}}, \bibinfo {author} {\bibfnamefont {S.}~\bibnamefont {Ravets}}, \bibinfo {author} {\bibfnamefont {S.}~\bibnamefont {De~L{\'e}s{\'e}leuc}}, \bibinfo {author} {\bibfnamefont {T.}~\bibnamefont {Macr{\`\i}}}, \bibinfo {author} {\bibfnamefont {T.}~\bibnamefont {Lahaye}},\ and\ \bibinfo {author} {\bibfnamefont {A.}~\bibnamefont {Browaeys}},\ }\bibfield  {title} {\bibinfo {title} {Tunable two-dimensional arrays of single {R}ydberg atoms for realizing quantum {I}sing models},\ }\href@noop {} {\bibfield  {journal} {\bibinfo  {journal} {Nature}\ }\textbf {\bibinfo {volume} {534}},\ \bibinfo {pages} {667} (\bibinfo {year} {2016})}\BibitemShut {NoStop}%
\bibitem [{\citenamefont {de~Mello}\ \emph {et~al.}(2019)\citenamefont {de~Mello}, \citenamefont {Sch{\"a}ffner}, \citenamefont {Werkmann}, \citenamefont {Preuschoff}, \citenamefont {Kohfahl}, \citenamefont {Schlosser},\ and\ \citenamefont {Birkl}}]{de2019defect}%
  \BibitemOpen
  \bibfield  {author} {\bibinfo {author} {\bibfnamefont {D.~O.}\ \bibnamefont {de~Mello}}, \bibinfo {author} {\bibfnamefont {D.}~\bibnamefont {Sch{\"a}ffner}}, \bibinfo {author} {\bibfnamefont {J.}~\bibnamefont {Werkmann}}, \bibinfo {author} {\bibfnamefont {T.}~\bibnamefont {Preuschoff}}, \bibinfo {author} {\bibfnamefont {L.}~\bibnamefont {Kohfahl}}, \bibinfo {author} {\bibfnamefont {M.}~\bibnamefont {Schlosser}},\ and\ \bibinfo {author} {\bibfnamefont {G.}~\bibnamefont {Birkl}},\ }\bibfield  {title} {\bibinfo {title} {Defect-free assembly of 2{D} clusters of more than 100 single-atom quantum systems},\ }\href@noop {} {\bibfield  {journal} {\bibinfo  {journal} {Phys. {R}ev. {L}ett.}\ }\textbf {\bibinfo {volume} {122}},\ \bibinfo {pages} {203601} (\bibinfo {year} {2019})}\BibitemShut {NoStop}%
\bibitem [{\citenamefont {Gambetta}\ \emph {et~al.}(2020)\citenamefont {Gambetta}, \citenamefont {Li}, \citenamefont {Schmidt-Kaler},\ and\ \citenamefont {Lesanovsky}}]{gambetta2020engineering}%
  \BibitemOpen
  \bibfield  {author} {\bibinfo {author} {\bibfnamefont {F.~M.}\ \bibnamefont {Gambetta}}, \bibinfo {author} {\bibfnamefont {W.}~\bibnamefont {Li}}, \bibinfo {author} {\bibfnamefont {F.}~\bibnamefont {Schmidt-Kaler}},\ and\ \bibinfo {author} {\bibfnamefont {I.}~\bibnamefont {Lesanovsky}},\ }\bibfield  {title} {\bibinfo {title} {Engineering nonbinary {R}ydberg interactions via phonons in an optical lattice},\ }\href@noop {} {\bibfield  {journal} {\bibinfo  {journal} {Phys. {R}ev. {L}ett.}\ }\textbf {\bibinfo {volume} {124}},\ \bibinfo {pages} {043402} (\bibinfo {year} {2020})}\BibitemShut {NoStop}%
\bibitem [{\citenamefont {Bernien}\ \emph {et~al.}(2017)\citenamefont {Bernien}, \citenamefont {Schwartz}, \citenamefont {Keesling}, \citenamefont {Levine}, \citenamefont {Omran}, \citenamefont {Pichler}, \citenamefont {Choi}, \citenamefont {Zibrov}, \citenamefont {Endres}, \citenamefont {Greiner} \emph {et~al.}}]{bernien2017probing}%
  \BibitemOpen
  \bibfield  {author} {\bibinfo {author} {\bibfnamefont {H.}~\bibnamefont {Bernien}}, \bibinfo {author} {\bibfnamefont {S.}~\bibnamefont {Schwartz}}, \bibinfo {author} {\bibfnamefont {A.}~\bibnamefont {Keesling}}, \bibinfo {author} {\bibfnamefont {H.}~\bibnamefont {Levine}}, \bibinfo {author} {\bibfnamefont {A.}~\bibnamefont {Omran}}, \bibinfo {author} {\bibfnamefont {H.}~\bibnamefont {Pichler}}, \bibinfo {author} {\bibfnamefont {S.}~\bibnamefont {Choi}}, \bibinfo {author} {\bibfnamefont {A.~S.}\ \bibnamefont {Zibrov}}, \bibinfo {author} {\bibfnamefont {M.}~\bibnamefont {Endres}}, \bibinfo {author} {\bibfnamefont {M.}~\bibnamefont {Greiner}}, \emph {et~al.},\ }\bibfield  {title} {\bibinfo {title} {Probing many-body dynamics on a 51-atom quantum simulator},\ }\href@noop {} {\bibfield  {journal} {\bibinfo  {journal} {Nature}\ }\textbf {\bibinfo {volume} {551}},\ \bibinfo {pages} {579} (\bibinfo {year} {2017})}\BibitemShut {NoStop}%
\bibitem [{\citenamefont {Kim}\ \emph {et~al.}(2018)\citenamefont {Kim}, \citenamefont {Park}, \citenamefont {Kim}, \citenamefont {Sim},\ and\ \citenamefont {Ahn}}]{kim2018detailed}%
  \BibitemOpen
  \bibfield  {author} {\bibinfo {author} {\bibfnamefont {H.}~\bibnamefont {Kim}}, \bibinfo {author} {\bibfnamefont {Y.}~\bibnamefont {Park}}, \bibinfo {author} {\bibfnamefont {K.}~\bibnamefont {Kim}}, \bibinfo {author} {\bibfnamefont {H.-S.}\ \bibnamefont {Sim}},\ and\ \bibinfo {author} {\bibfnamefont {J.}~\bibnamefont {Ahn}},\ }\bibfield  {title} {\bibinfo {title} {Detailed balance of thermalization dynamics in {R}ydberg-atom quantum simulators},\ }\href@noop {} {\bibfield  {journal} {\bibinfo  {journal} {Phys. {R}ev. {L}ett.}\ }\textbf {\bibinfo {volume} {120}},\ \bibinfo {pages} {180502} (\bibinfo {year} {2018})}\BibitemShut {NoStop}%
\bibitem [{\citenamefont {Lienhard}\ \emph {et~al.}(2018)\citenamefont {Lienhard}, \citenamefont {de~L{\'e}s{\'e}leuc}, \citenamefont {Barredo}, \citenamefont {Lahaye}, \citenamefont {Browaeys}, \citenamefont {Schuler}, \citenamefont {Henry},\ and\ \citenamefont {L{\"a}uchli}}]{lienhard2018observing}%
  \BibitemOpen
  \bibfield  {author} {\bibinfo {author} {\bibfnamefont {V.}~\bibnamefont {Lienhard}}, \bibinfo {author} {\bibfnamefont {S.}~\bibnamefont {de~L{\'e}s{\'e}leuc}}, \bibinfo {author} {\bibfnamefont {D.}~\bibnamefont {Barredo}}, \bibinfo {author} {\bibfnamefont {T.}~\bibnamefont {Lahaye}}, \bibinfo {author} {\bibfnamefont {A.}~\bibnamefont {Browaeys}}, \bibinfo {author} {\bibfnamefont {M.}~\bibnamefont {Schuler}}, \bibinfo {author} {\bibfnamefont {L.-P.}\ \bibnamefont {Henry}},\ and\ \bibinfo {author} {\bibfnamefont {A.~M.}\ \bibnamefont {L{\"a}uchli}},\ }\bibfield  {title} {\bibinfo {title} {Observing the space-and time-dependent growth of correlations in dynamically tuned synthetic {I}sing models with antiferromagnetic interactions},\ }\href@noop {} {\bibfield  {journal} {\bibinfo  {journal} {Phys. {R}ev. {X}}\ }\textbf {\bibinfo {volume} {8}},\ \bibinfo {pages} {021070} (\bibinfo {year} {2018})}\BibitemShut {NoStop}%
\bibitem [{\citenamefont {De~L{\'e}s{\'e}leuc}\ \emph {et~al.}(2018)\citenamefont {De~L{\'e}s{\'e}leuc}, \citenamefont {Weber}, \citenamefont {Lienhard}, \citenamefont {Barredo}, \citenamefont {B{\"u}chler}, \citenamefont {Lahaye},\ and\ \citenamefont {Browaeys}}]{de2018accurate}%
  \BibitemOpen
  \bibfield  {author} {\bibinfo {author} {\bibfnamefont {S.}~\bibnamefont {De~L{\'e}s{\'e}leuc}}, \bibinfo {author} {\bibfnamefont {S.}~\bibnamefont {Weber}}, \bibinfo {author} {\bibfnamefont {V.}~\bibnamefont {Lienhard}}, \bibinfo {author} {\bibfnamefont {D.}~\bibnamefont {Barredo}}, \bibinfo {author} {\bibfnamefont {H.~P.}\ \bibnamefont {B{\"u}chler}}, \bibinfo {author} {\bibfnamefont {T.}~\bibnamefont {Lahaye}},\ and\ \bibinfo {author} {\bibfnamefont {A.}~\bibnamefont {Browaeys}},\ }\bibfield  {title} {\bibinfo {title} {Accurate mapping of multilevel {R}ydberg atoms on interacting spin-1/2 particles for the quantum simulation of {I}sing models},\ }\href@noop {} {\bibfield  {journal} {\bibinfo  {journal} {Phys. {R}ev. {L}ett.}\ }\textbf {\bibinfo {volume} {120}},\ \bibinfo {pages} {113602} (\bibinfo {year} {2018})}\BibitemShut {NoStop}%
\bibitem [{\citenamefont {Scholl}\ \emph {et~al.}(2021)\citenamefont {Scholl}, \citenamefont {Schuler}, \citenamefont {Williams}, \citenamefont {Eberharter}, \citenamefont {Barredo}, \citenamefont {Schymik}, \citenamefont {Lienhard}, \citenamefont {Henry}, \citenamefont {Lang}, \citenamefont {Lahaye} \emph {et~al.}}]{scholl2020programmable}%
  \BibitemOpen
  \bibfield  {author} {\bibinfo {author} {\bibfnamefont {P.}~\bibnamefont {Scholl}}, \bibinfo {author} {\bibfnamefont {M.}~\bibnamefont {Schuler}}, \bibinfo {author} {\bibfnamefont {H.~J.}\ \bibnamefont {Williams}}, \bibinfo {author} {\bibfnamefont {A.~A.}\ \bibnamefont {Eberharter}}, \bibinfo {author} {\bibfnamefont {D.}~\bibnamefont {Barredo}}, \bibinfo {author} {\bibfnamefont {K.-N.}\ \bibnamefont {Schymik}}, \bibinfo {author} {\bibfnamefont {V.}~\bibnamefont {Lienhard}}, \bibinfo {author} {\bibfnamefont {L.-P.}\ \bibnamefont {Henry}}, \bibinfo {author} {\bibfnamefont {T.~C.}\ \bibnamefont {Lang}}, \bibinfo {author} {\bibfnamefont {T.}~\bibnamefont {Lahaye}}, \emph {et~al.},\ }\bibfield  {title} {\bibinfo {title} {Quantum simulation of 2{D} antiferromagnets with hundreds of {R}ydberg atoms},\ }\href@noop {} {\bibfield  {journal} {\bibinfo  {journal} {Nature}\ }\textbf {\bibinfo {volume} {595}},\ \bibinfo {pages} {233} (\bibinfo {year} {2021})}\BibitemShut {NoStop}%
\bibitem [{\citenamefont {Browaeys}\ and\ \citenamefont {Lahaye}(2020)}]{browaeys2020many}%
  \BibitemOpen
  \bibfield  {author} {\bibinfo {author} {\bibfnamefont {A.}~\bibnamefont {Browaeys}}\ and\ \bibinfo {author} {\bibfnamefont {T.}~\bibnamefont {Lahaye}},\ }\bibfield  {title} {\bibinfo {title} {Many-body physics with individually controlled {R}ydberg atoms},\ }\href@noop {} {\bibfield  {journal} {\bibinfo  {journal} {Nat. {P}hys.}\ }\textbf {\bibinfo {volume} {16}},\ \bibinfo {pages} {132} (\bibinfo {year} {2020})}\BibitemShut {NoStop}%
\bibitem [{\citenamefont {Ebadi}\ \emph {et~al.}(2021)\citenamefont {Ebadi}, \citenamefont {Wang}, \citenamefont {Levine}, \citenamefont {Keesling}, \citenamefont {Semeghini}, \citenamefont {Omran}, \citenamefont {Bluvstein}, \citenamefont {Samajdar}, \citenamefont {Pichler}, \citenamefont {Ho} \emph {et~al.}}]{ebadi2021quantum}%
  \BibitemOpen
  \bibfield  {author} {\bibinfo {author} {\bibfnamefont {S.}~\bibnamefont {Ebadi}}, \bibinfo {author} {\bibfnamefont {T.~T.}\ \bibnamefont {Wang}}, \bibinfo {author} {\bibfnamefont {H.}~\bibnamefont {Levine}}, \bibinfo {author} {\bibfnamefont {A.}~\bibnamefont {Keesling}}, \bibinfo {author} {\bibfnamefont {G.}~\bibnamefont {Semeghini}}, \bibinfo {author} {\bibfnamefont {A.}~\bibnamefont {Omran}}, \bibinfo {author} {\bibfnamefont {D.}~\bibnamefont {Bluvstein}}, \bibinfo {author} {\bibfnamefont {R.}~\bibnamefont {Samajdar}}, \bibinfo {author} {\bibfnamefont {H.}~\bibnamefont {Pichler}}, \bibinfo {author} {\bibfnamefont {W.~W.}\ \bibnamefont {Ho}}, \emph {et~al.},\ }\bibfield  {title} {\bibinfo {title} {Quantum phases of matter on a 256-atom programmable quantum simulator},\ }\href@noop {} {\bibfield  {journal} {\bibinfo  {journal} {Nature}\ }\textbf {\bibinfo {volume} {595}},\ \bibinfo {pages} {227} (\bibinfo {year} {2021})}\BibitemShut {NoStop}%
\bibitem [{\citenamefont {Semeghini}\ \emph {et~al.}(2021)\citenamefont {Semeghini}, \citenamefont {Levine}, \citenamefont {Keesling}, \citenamefont {Ebadi}, \citenamefont {Wang}, \citenamefont {Bluvstein}, \citenamefont {Verresen}, \citenamefont {Pichler}, \citenamefont {Kalinowski}, \citenamefont {Samajdar}, \citenamefont {Omran}, \citenamefont {Sachdev}, \citenamefont {Vishwanath}, \citenamefont {Greiner}, \citenamefont {Vuletić},\ and\ \citenamefont {Lukin}}]{semeghini2021probing}%
  \BibitemOpen
  \bibfield  {author} {\bibinfo {author} {\bibfnamefont {G.}~\bibnamefont {Semeghini}}, \bibinfo {author} {\bibfnamefont {H.}~\bibnamefont {Levine}}, \bibinfo {author} {\bibfnamefont {A.}~\bibnamefont {Keesling}}, \bibinfo {author} {\bibfnamefont {S.}~\bibnamefont {Ebadi}}, \bibinfo {author} {\bibfnamefont {T.~T.}\ \bibnamefont {Wang}}, \bibinfo {author} {\bibfnamefont {D.}~\bibnamefont {Bluvstein}}, \bibinfo {author} {\bibfnamefont {R.}~\bibnamefont {Verresen}}, \bibinfo {author} {\bibfnamefont {H.}~\bibnamefont {Pichler}}, \bibinfo {author} {\bibfnamefont {M.}~\bibnamefont {Kalinowski}}, \bibinfo {author} {\bibfnamefont {R.}~\bibnamefont {Samajdar}}, \bibinfo {author} {\bibfnamefont {A.}~\bibnamefont {Omran}}, \bibinfo {author} {\bibfnamefont {S.}~\bibnamefont {Sachdev}}, \bibinfo {author} {\bibfnamefont {A.}~\bibnamefont {Vishwanath}}, \bibinfo {author} {\bibfnamefont {M.}~\bibnamefont {Greiner}}, \bibinfo {author} {\bibfnamefont {V.}~\bibnamefont {Vuletić}},\ and\ \bibinfo {author} {\bibfnamefont
  {M.~D.}\ \bibnamefont {Lukin}},\ }\bibfield  {title} {\bibinfo {title} {Probing topological spin liquids on a programmable quantum simulator},\ }\href@noop {} {\bibfield  {journal} {\bibinfo  {journal} {Science}\ }\textbf {\bibinfo {volume} {374}},\ \bibinfo {pages} {1242} (\bibinfo {year} {2021})}\BibitemShut {NoStop}%
\bibitem [{\citenamefont {Song}\ \emph {et~al.}(2021)\citenamefont {Song}, \citenamefont {Kim}, \citenamefont {Hwang}, \citenamefont {Lee},\ and\ \citenamefont {Ahn}}]{song2020quantum}%
  \BibitemOpen
  \bibfield  {author} {\bibinfo {author} {\bibfnamefont {Y.}~\bibnamefont {Song}}, \bibinfo {author} {\bibfnamefont {M.}~\bibnamefont {Kim}}, \bibinfo {author} {\bibfnamefont {H.}~\bibnamefont {Hwang}}, \bibinfo {author} {\bibfnamefont {W.}~\bibnamefont {Lee}},\ and\ \bibinfo {author} {\bibfnamefont {J.}~\bibnamefont {Ahn}},\ }\bibfield  {title} {\bibinfo {title} {Quantum simulation of {C}ayley-tree {I}sing {H}amiltonians with three-dimensional {R}ydberg atoms},\ }\href@noop {} {\bibfield  {journal} {\bibinfo  {journal} {Phys. {R}ev. {R}es.}\ }\textbf {\bibinfo {volume} {3}},\ \bibinfo {pages} {013286} (\bibinfo {year} {2021})}\BibitemShut {NoStop}%
\bibitem [{\citenamefont {Dauphin}\ \emph {et~al.}(2016)\citenamefont {Dauphin}, \citenamefont {M{\"u}ller},\ and\ \citenamefont {Martin-Delgado}}]{dauphin2016quantum}%
  \BibitemOpen
  \bibfield  {author} {\bibinfo {author} {\bibfnamefont {A.}~\bibnamefont {Dauphin}}, \bibinfo {author} {\bibfnamefont {M.}~\bibnamefont {M{\"u}ller}},\ and\ \bibinfo {author} {\bibfnamefont {M.~A.}\ \bibnamefont {Martin-Delgado}},\ }\bibfield  {title} {\bibinfo {title} {Quantum simulation of a topological {M}ott insulator with {R}ydberg atoms in a {L}ieb lattice},\ }\href@noop {} {\bibfield  {journal} {\bibinfo  {journal} {Phys. {R}ev. {A}}\ }\textbf {\bibinfo {volume} {93}},\ \bibinfo {pages} {043611} (\bibinfo {year} {2016})}\BibitemShut {NoStop}%
\bibitem [{\citenamefont {Grusdt}\ \emph {et~al.}(2018)\citenamefont {Grusdt}, \citenamefont {K{\'a}nasz-Nagy}, \citenamefont {Bohrdt}, \citenamefont {Chiu}, \citenamefont {Ji}, \citenamefont {Greiner}, \citenamefont {Greif},\ and\ \citenamefont {Demler}}]{grusdt2018parton}%
  \BibitemOpen
  \bibfield  {author} {\bibinfo {author} {\bibfnamefont {F.}~\bibnamefont {Grusdt}}, \bibinfo {author} {\bibfnamefont {M.}~\bibnamefont {K{\'a}nasz-Nagy}}, \bibinfo {author} {\bibfnamefont {A.}~\bibnamefont {Bohrdt}}, \bibinfo {author} {\bibfnamefont {C.~S.}\ \bibnamefont {Chiu}}, \bibinfo {author} {\bibfnamefont {G.}~\bibnamefont {Ji}}, \bibinfo {author} {\bibfnamefont {M.}~\bibnamefont {Greiner}}, \bibinfo {author} {\bibfnamefont {D.}~\bibnamefont {Greif}},\ and\ \bibinfo {author} {\bibfnamefont {E.}~\bibnamefont {Demler}},\ }\bibfield  {title} {\bibinfo {title} {Parton theory of magnetic polarons: {M}esonic resonances and signatures in dynamics},\ }\href@noop {} {\bibfield  {journal} {\bibinfo  {journal} {Phys. {R}ev. {X}}\ }\textbf {\bibinfo {volume} {8}},\ \bibinfo {pages} {011046} (\bibinfo {year} {2018})}\BibitemShut {NoStop}%
\bibitem [{\citenamefont {Guardado-Sanchez}\ \emph {et~al.}(2018)\citenamefont {Guardado-Sanchez}, \citenamefont {Brown}, \citenamefont {Mitra}, \citenamefont {Devakul}, \citenamefont {Huse}, \citenamefont {Schau{\ss}},\ and\ \citenamefont {Bakr}}]{guardado2018probing}%
  \BibitemOpen
  \bibfield  {author} {\bibinfo {author} {\bibfnamefont {E.}~\bibnamefont {Guardado-Sanchez}}, \bibinfo {author} {\bibfnamefont {P.~T.}\ \bibnamefont {Brown}}, \bibinfo {author} {\bibfnamefont {D.}~\bibnamefont {Mitra}}, \bibinfo {author} {\bibfnamefont {T.}~\bibnamefont {Devakul}}, \bibinfo {author} {\bibfnamefont {D.~A.}\ \bibnamefont {Huse}}, \bibinfo {author} {\bibfnamefont {P.}~\bibnamefont {Schau{\ss}}},\ and\ \bibinfo {author} {\bibfnamefont {W.~S.}\ \bibnamefont {Bakr}},\ }\bibfield  {title} {\bibinfo {title} {Probing the quench dynamics of antiferromagnetic correlations in a 2{D} quantum {I}sing spin system},\ }\href@noop {} {\bibfield  {journal} {\bibinfo  {journal} {Phys. {R}ev. {X}}\ }\textbf {\bibinfo {volume} {8}},\ \bibinfo {pages} {021069} (\bibinfo {year} {2018})}\BibitemShut {NoStop}%
\bibitem [{\citenamefont {Li}\ \emph {et~al.}(2013)\citenamefont {Li}, \citenamefont {Ates},\ and\ \citenamefont {Lesanovsky}}]{li2013nonadiabatic}%
  \BibitemOpen
  \bibfield  {author} {\bibinfo {author} {\bibfnamefont {W.}~\bibnamefont {Li}}, \bibinfo {author} {\bibfnamefont {C.}~\bibnamefont {Ates}},\ and\ \bibinfo {author} {\bibfnamefont {I.}~\bibnamefont {Lesanovsky}},\ }\bibfield  {title} {\bibinfo {title} {Nonadiabatic motional effects and dissipative blockade for {R}ydberg atoms excited from optical lattices or microtraps},\ }\href@noop {} {\bibfield  {journal} {\bibinfo  {journal} {Phys. {R}ev. {L}ett.}\ }\textbf {\bibinfo {volume} {110}},\ \bibinfo {pages} {213005} (\bibinfo {year} {2013})}\BibitemShut {NoStop}%
\bibitem [{\citenamefont {Macri}\ and\ \citenamefont {Pohl}(2014)}]{macri2014rydberg}%
  \BibitemOpen
  \bibfield  {author} {\bibinfo {author} {\bibfnamefont {T.}~\bibnamefont {Macri}}\ and\ \bibinfo {author} {\bibfnamefont {T.}~\bibnamefont {Pohl}},\ }\bibfield  {title} {\bibinfo {title} {Rydberg dressing of atoms in optical lattices},\ }\href@noop {} {\bibfield  {journal} {\bibinfo  {journal} {Phys. {R}ev. A}\ }\textbf {\bibinfo {volume} {89}},\ \bibinfo {pages} {011402} (\bibinfo {year} {2014})}\BibitemShut {NoStop}%
\bibitem [{\citenamefont {Festa}\ \emph {et~al.}(2022)\citenamefont {Festa}, \citenamefont {Lorenz}, \citenamefont {Steinert}, \citenamefont {Chen}, \citenamefont {Osterholz}, \citenamefont {Eberhard},\ and\ \citenamefont {Gross}}]{festa2021motion}%
  \BibitemOpen
  \bibfield  {author} {\bibinfo {author} {\bibfnamefont {L.}~\bibnamefont {Festa}}, \bibinfo {author} {\bibfnamefont {N.}~\bibnamefont {Lorenz}}, \bibinfo {author} {\bibfnamefont {L.-M.}\ \bibnamefont {Steinert}}, \bibinfo {author} {\bibfnamefont {Z.}~\bibnamefont {Chen}}, \bibinfo {author} {\bibfnamefont {P.}~\bibnamefont {Osterholz}}, \bibinfo {author} {\bibfnamefont {R.}~\bibnamefont {Eberhard}},\ and\ \bibinfo {author} {\bibfnamefont {C.}~\bibnamefont {Gross}},\ }\bibfield  {title} {\bibinfo {title} {Blackbody-radiation-induced facilitated excitation of {R}ydberg atoms in optical tweezers},\ }\href@noop {} {\bibfield  {journal} {\bibinfo  {journal} {Phys. {R}ev. {A}}\ }\textbf {\bibinfo {volume} {105}},\ \bibinfo {pages} {013109} (\bibinfo {year} {2022})}\BibitemShut {NoStop}%
\bibitem [{\citenamefont {Robicheaux}\ \emph {et~al.}(2021)\citenamefont {Robicheaux}, \citenamefont {Graham},\ and\ \citenamefont {Saffman}}]{robicheaux2021photon}%
  \BibitemOpen
  \bibfield  {author} {\bibinfo {author} {\bibfnamefont {F.}~\bibnamefont {Robicheaux}}, \bibinfo {author} {\bibfnamefont {T.}~\bibnamefont {Graham}},\ and\ \bibinfo {author} {\bibfnamefont {M.}~\bibnamefont {Saffman}},\ }\bibfield  {title} {\bibinfo {title} {Photon-recoil and laser-focusing limits to {R}ydberg gate fidelity},\ }\href@noop {} {\bibfield  {journal} {\bibinfo  {journal} {Phys. Rev. A}\ }\textbf {\bibinfo {volume} {103}},\ \bibinfo {pages} {022424} (\bibinfo {year} {2021})}\BibitemShut {NoStop}%
\bibitem [{\citenamefont {Magoni}\ \emph {et~al.}(2022)\citenamefont {Magoni}, \citenamefont {Mazza},\ and\ \citenamefont {Lesanovsky}}]{magoni2022phonon}%
  \BibitemOpen
  \bibfield  {author} {\bibinfo {author} {\bibfnamefont {M.}~\bibnamefont {Magoni}}, \bibinfo {author} {\bibfnamefont {P.}~\bibnamefont {Mazza}},\ and\ \bibinfo {author} {\bibfnamefont {I.}~\bibnamefont {Lesanovsky}},\ }\bibfield  {title} {\bibinfo {title} {Phonon dressing of a facilitated one-dimensional {R}ydberg lattice gas},\ }\href@noop {} {\bibfield  {journal} {\bibinfo  {journal} {SciPost {P}hys. {C}ore}\ }\textbf {\bibinfo {volume} {5}},\ \bibinfo {pages} {041} (\bibinfo {year} {2022})}\BibitemShut {NoStop}%
\bibitem [{\citenamefont {Dudin}\ and\ \citenamefont {Kuzmich}(2012)}]{dudin2012strongly}%
  \BibitemOpen
  \bibfield  {author} {\bibinfo {author} {\bibfnamefont {Y.}~\bibnamefont {Dudin}}\ and\ \bibinfo {author} {\bibfnamefont {A.}~\bibnamefont {Kuzmich}},\ }\bibfield  {title} {\bibinfo {title} {Strongly interacting {R}ydberg excitations of a cold atomic gas},\ }\href@noop {} {\bibfield  {journal} {\bibinfo  {journal} {Science}\ }\textbf {\bibinfo {volume} {336}},\ \bibinfo {pages} {887} (\bibinfo {year} {2012})}\BibitemShut {NoStop}%
\bibitem [{\citenamefont {Baur}\ \emph {et~al.}(2014)\citenamefont {Baur}, \citenamefont {Tiarks}, \citenamefont {Rempe},\ and\ \citenamefont {D{\"u}rr}}]{baur2014single}%
  \BibitemOpen
  \bibfield  {author} {\bibinfo {author} {\bibfnamefont {S.}~\bibnamefont {Baur}}, \bibinfo {author} {\bibfnamefont {D.}~\bibnamefont {Tiarks}}, \bibinfo {author} {\bibfnamefont {G.}~\bibnamefont {Rempe}},\ and\ \bibinfo {author} {\bibfnamefont {S.}~\bibnamefont {D{\"u}rr}},\ }\bibfield  {title} {\bibinfo {title} {Single-photon switch based on {R}ydberg blockade},\ }\href@noop {} {\bibfield  {journal} {\bibinfo  {journal} {Phys. {R}ev. {L}ett.}\ }\textbf {\bibinfo {volume} {112}},\ \bibinfo {pages} {073901} (\bibinfo {year} {2014})}\BibitemShut {NoStop}%
\bibitem [{\citenamefont {M{\"u}ller}\ \emph {et~al.}(2014)\citenamefont {M{\"u}ller}, \citenamefont {Murphy}, \citenamefont {Montangero}, \citenamefont {Calarco}, \citenamefont {Grangier},\ and\ \citenamefont {Browaeys}}]{muller2014implementation}%
  \BibitemOpen
  \bibfield  {author} {\bibinfo {author} {\bibfnamefont {M.~M.}\ \bibnamefont {M{\"u}ller}}, \bibinfo {author} {\bibfnamefont {M.}~\bibnamefont {Murphy}}, \bibinfo {author} {\bibfnamefont {S.}~\bibnamefont {Montangero}}, \bibinfo {author} {\bibfnamefont {T.}~\bibnamefont {Calarco}}, \bibinfo {author} {\bibfnamefont {P.}~\bibnamefont {Grangier}},\ and\ \bibinfo {author} {\bibfnamefont {A.}~\bibnamefont {Browaeys}},\ }\bibfield  {title} {\bibinfo {title} {Implementation of an experimentally feasible controlled-phase gate on two blockaded {R}ydberg atoms},\ }\href@noop {} {\bibfield  {journal} {\bibinfo  {journal} {Phys. {R}ev. {A}}\ }\textbf {\bibinfo {volume} {89}},\ \bibinfo {pages} {032334} (\bibinfo {year} {2014})}\BibitemShut {NoStop}%
\bibitem [{\citenamefont {Han}\ \emph {et~al.}(2016)\citenamefont {Han}, \citenamefont {Vogt},\ and\ \citenamefont {Li}}]{han2016spectral}%
  \BibitemOpen
  \bibfield  {author} {\bibinfo {author} {\bibfnamefont {J.}~\bibnamefont {Han}}, \bibinfo {author} {\bibfnamefont {T.}~\bibnamefont {Vogt}},\ and\ \bibinfo {author} {\bibfnamefont {W.}~\bibnamefont {Li}},\ }\bibfield  {title} {\bibinfo {title} {Spectral shift and dephasing of electromagnetically induced transparency in an interacting {R}ydberg gas},\ }\href@noop {} {\bibfield  {journal} {\bibinfo  {journal} {Phys. {R}ev. A}\ }\textbf {\bibinfo {volume} {94}},\ \bibinfo {pages} {043806} (\bibinfo {year} {2016})}\BibitemShut {NoStop}%
\bibitem [{\citenamefont {Marcuzzi}\ \emph {et~al.}(2017)\citenamefont {Marcuzzi}, \citenamefont {Min{\'a}{\v{r}}}, \citenamefont {Barredo}, \citenamefont {De~L{\'e}s{\'e}leuc}, \citenamefont {Labuhn}, \citenamefont {Lahaye}, \citenamefont {Browaeys}, \citenamefont {Levi},\ and\ \citenamefont {Lesanovsky}}]{marcuzzi2017facilitation}%
  \BibitemOpen
  \bibfield  {author} {\bibinfo {author} {\bibfnamefont {M.}~\bibnamefont {Marcuzzi}}, \bibinfo {author} {\bibfnamefont {J.}~\bibnamefont {Min{\'a}{\v{r}}}}, \bibinfo {author} {\bibfnamefont {D.}~\bibnamefont {Barredo}}, \bibinfo {author} {\bibfnamefont {S.}~\bibnamefont {De~L{\'e}s{\'e}leuc}}, \bibinfo {author} {\bibfnamefont {H.}~\bibnamefont {Labuhn}}, \bibinfo {author} {\bibfnamefont {T.}~\bibnamefont {Lahaye}}, \bibinfo {author} {\bibfnamefont {A.}~\bibnamefont {Browaeys}}, \bibinfo {author} {\bibfnamefont {E.}~\bibnamefont {Levi}},\ and\ \bibinfo {author} {\bibfnamefont {I.}~\bibnamefont {Lesanovsky}},\ }\bibfield  {title} {\bibinfo {title} {Facilitation dynamics and localization phenomena in {R}ydberg lattice gases with position disorder},\ }\href@noop {} {\bibfield  {journal} {\bibinfo  {journal} {Phys. {R}ev. {L}ett.}\ }\textbf {\bibinfo {volume} {118}},\ \bibinfo {pages} {063606} (\bibinfo {year} {2017})}\BibitemShut {NoStop}%
\bibitem [{\citenamefont {Schachenmayer}\ \emph {et~al.}(2015{\natexlab{a}})\citenamefont {Schachenmayer}, \citenamefont {Pikovski},\ and\ \citenamefont {Rey}}]{schachenmayer2015many}%
  \BibitemOpen
  \bibfield  {author} {\bibinfo {author} {\bibfnamefont {J.}~\bibnamefont {Schachenmayer}}, \bibinfo {author} {\bibfnamefont {A.}~\bibnamefont {Pikovski}},\ and\ \bibinfo {author} {\bibfnamefont {A.~M.}\ \bibnamefont {Rey}},\ }\bibfield  {title} {\bibinfo {title} {Many-body quantum spin dynamics with {M}onte {C}arlo trajectories on a discrete phase space},\ }\href@noop {} {\bibfield  {journal} {\bibinfo  {journal} {Phys. {R}ev. {X}}\ }\textbf {\bibinfo {volume} {5}},\ \bibinfo {pages} {011022} (\bibinfo {year} {2015}{\natexlab{a}})}\BibitemShut {NoStop}%
\bibitem [{\citenamefont {Khasseh}\ \emph {et~al.}(2020)\citenamefont {Khasseh}, \citenamefont {Russomanno}, \citenamefont {Schmitt}, \citenamefont {Heyl},\ and\ \citenamefont {Fazio}}]{khasseh2020discrete}%
  \BibitemOpen
  \bibfield  {author} {\bibinfo {author} {\bibfnamefont {R.}~\bibnamefont {Khasseh}}, \bibinfo {author} {\bibfnamefont {A.}~\bibnamefont {Russomanno}}, \bibinfo {author} {\bibfnamefont {M.}~\bibnamefont {Schmitt}}, \bibinfo {author} {\bibfnamefont {M.}~\bibnamefont {Heyl}},\ and\ \bibinfo {author} {\bibfnamefont {R.}~\bibnamefont {Fazio}},\ }\bibfield  {title} {\bibinfo {title} {Discrete truncated {W}igner approach to dynamical phase transitions in {I}sing models after a quantum quench},\ }\href@noop {} {\bibfield  {journal} {\bibinfo  {journal} {Phys. {R}ev. {B}}\ }\textbf {\bibinfo {volume} {102}},\ \bibinfo {pages} {014303} (\bibinfo {year} {2020})}\BibitemShut {NoStop}%
\bibitem [{\citenamefont {Kunimi}\ \emph {et~al.}(2021)\citenamefont {Kunimi}, \citenamefont {Nagao}, \citenamefont {Goto},\ and\ \citenamefont {Danshita}}]{kunimi2021performance}%
  \BibitemOpen
  \bibfield  {author} {\bibinfo {author} {\bibfnamefont {M.}~\bibnamefont {Kunimi}}, \bibinfo {author} {\bibfnamefont {K.}~\bibnamefont {Nagao}}, \bibinfo {author} {\bibfnamefont {S.}~\bibnamefont {Goto}},\ and\ \bibinfo {author} {\bibfnamefont {I.}~\bibnamefont {Danshita}},\ }\bibfield  {title} {\bibinfo {title} {Performance evaluation of the discrete truncated {W}igner approximation for quench dynamics of quantum spin systems with long-range interactions},\ }\href@noop {} {\bibfield  {journal} {\bibinfo  {journal} {Phys. {R}ev. {R}es.}\ }\textbf {\bibinfo {volume} {3}},\ \bibinfo {pages} {013060} (\bibinfo {year} {2021})}\BibitemShut {NoStop}%
\bibitem [{\citenamefont {Czischek}\ \emph {et~al.}(2018)\citenamefont {Czischek}, \citenamefont {G{\"a}rttner}, \citenamefont {Oberthaler}, \citenamefont {Kastner},\ and\ \citenamefont {Gasenzer}}]{czischek2018quenches}%
  \BibitemOpen
  \bibfield  {author} {\bibinfo {author} {\bibfnamefont {S.}~\bibnamefont {Czischek}}, \bibinfo {author} {\bibfnamefont {M.}~\bibnamefont {G{\"a}rttner}}, \bibinfo {author} {\bibfnamefont {M.}~\bibnamefont {Oberthaler}}, \bibinfo {author} {\bibfnamefont {M.}~\bibnamefont {Kastner}},\ and\ \bibinfo {author} {\bibfnamefont {T.}~\bibnamefont {Gasenzer}},\ }\bibfield  {title} {\bibinfo {title} {Quenches near criticality of the quantum {I}sing chain—power and limitations of the discrete truncated {W}igner approximation},\ }\href@noop {} {\bibfield  {journal} {\bibinfo  {journal} {Quantum {S}ci. {T}echnol.}\ }\textbf {\bibinfo {volume} {4}},\ \bibinfo {pages} {014006} (\bibinfo {year} {2018})}\BibitemShut {NoStop}%
\bibitem [{\citenamefont {Polkovnikov}(2010)}]{polkovnikov2010phase}%
  \BibitemOpen
  \bibfield  {author} {\bibinfo {author} {\bibfnamefont {A.}~\bibnamefont {Polkovnikov}},\ }\bibfield  {title} {\bibinfo {title} {Phase space representation of quantum dynamics},\ }\href@noop {} {\bibfield  {journal} {\bibinfo  {journal} {Ann. {P}hys.}\ }\textbf {\bibinfo {volume} {325}},\ \bibinfo {pages} {1790} (\bibinfo {year} {2010})}\BibitemShut {NoStop}%
\bibitem [{\citenamefont {Pucci}\ \emph {et~al.}(2016)\citenamefont {Pucci}, \citenamefont {Roy},\ and\ \citenamefont {Kastner}}]{pucci2016simulation}%
  \BibitemOpen
  \bibfield  {author} {\bibinfo {author} {\bibfnamefont {L.}~\bibnamefont {Pucci}}, \bibinfo {author} {\bibfnamefont {A.}~\bibnamefont {Roy}},\ and\ \bibinfo {author} {\bibfnamefont {M.}~\bibnamefont {Kastner}},\ }\bibfield  {title} {\bibinfo {title} {Simulation of quantum spin dynamics by phase space sampling of {B}ogoliubov-{B}orn-{G}reen-{K}irkwood-{Y}von trajectories},\ }\href@noop {} {\bibfield  {journal} {\bibinfo  {journal} {Phys. {R}ev. B}\ }\textbf {\bibinfo {volume} {93}},\ \bibinfo {pages} {174302} (\bibinfo {year} {2016})}\BibitemShut {NoStop}%
\bibitem [{\citenamefont {Acevedo}\ \emph {et~al.}(2017)\citenamefont {Acevedo}, \citenamefont {Safavi-Naini}, \citenamefont {Schachenmayer}, \citenamefont {Wall}, \citenamefont {Nandkishore},\ and\ \citenamefont {Rey}}]{acevedo2017exploring}%
  \BibitemOpen
  \bibfield  {author} {\bibinfo {author} {\bibfnamefont {O.}~\bibnamefont {Acevedo}}, \bibinfo {author} {\bibfnamefont {A.}~\bibnamefont {Safavi-Naini}}, \bibinfo {author} {\bibfnamefont {J.}~\bibnamefont {Schachenmayer}}, \bibinfo {author} {\bibfnamefont {M.}~\bibnamefont {Wall}}, \bibinfo {author} {\bibfnamefont {R.}~\bibnamefont {Nandkishore}},\ and\ \bibinfo {author} {\bibfnamefont {A.}~\bibnamefont {Rey}},\ }\bibfield  {title} {\bibinfo {title} {Exploring many-body localization and thermalization using semiclassical methods},\ }\href@noop {} {\bibfield  {journal} {\bibinfo  {journal} {Phys. {R}ev. A}\ }\textbf {\bibinfo {volume} {96}},\ \bibinfo {pages} {033604} (\bibinfo {year} {2017})}\BibitemShut {NoStop}%
\bibitem [{\citenamefont {Sundar}\ \emph {et~al.}(2019)\citenamefont {Sundar}, \citenamefont {Wang},\ and\ \citenamefont {Hazzard}}]{sundar2019analysis}%
  \BibitemOpen
  \bibfield  {author} {\bibinfo {author} {\bibfnamefont {B.}~\bibnamefont {Sundar}}, \bibinfo {author} {\bibfnamefont {K.~C.}\ \bibnamefont {Wang}},\ and\ \bibinfo {author} {\bibfnamefont {K.~R.~A.}\ \bibnamefont {Hazzard}},\ }\bibfield  {title} {\bibinfo {title} {Analysis of continuous and discrete {W}igner approximations for spin dynamics},\ }\href@noop {} {\bibfield  {journal} {\bibinfo  {journal} {Phys. {R}ev. {A}}\ }\textbf {\bibinfo {volume} {99}},\ \bibinfo {pages} {043627} (\bibinfo {year} {2019})}\BibitemShut {NoStop}%
\bibitem [{\citenamefont {Schachenmayer}\ \emph {et~al.}(2015{\natexlab{b}})\citenamefont {Schachenmayer}, \citenamefont {Pikovski},\ and\ \citenamefont {Rey}}]{schachenmayer2015dynamics}%
  \BibitemOpen
  \bibfield  {author} {\bibinfo {author} {\bibfnamefont {J.}~\bibnamefont {Schachenmayer}}, \bibinfo {author} {\bibfnamefont {A.}~\bibnamefont {Pikovski}},\ and\ \bibinfo {author} {\bibfnamefont {A.~M.}\ \bibnamefont {Rey}},\ }\bibfield  {title} {\bibinfo {title} {Dynamics of correlations in two-dimensional quantum spin models with long-range interactions: a phase-space {M}onte-{C}arlo study},\ }\href@noop {} {\bibfield  {journal} {\bibinfo  {journal} {New {J}. {P}hys.}\ }\textbf {\bibinfo {volume} {17}},\ \bibinfo {pages} {065009} (\bibinfo {year} {2015}{\natexlab{b}})}\BibitemShut {NoStop}%
\bibitem [{\citenamefont {Signoles}\ \emph {et~al.}(2021)\citenamefont {Signoles}, \citenamefont {Franz}, \citenamefont {Alves}, \citenamefont {G{\"a}rttner}, \citenamefont {Whitlock}, \citenamefont {Z{\"u}rn},\ and\ \citenamefont {Weidem{\"u}ller}}]{signoles2021glassy}%
  \BibitemOpen
  \bibfield  {author} {\bibinfo {author} {\bibfnamefont {A.}~\bibnamefont {Signoles}}, \bibinfo {author} {\bibfnamefont {T.}~\bibnamefont {Franz}}, \bibinfo {author} {\bibfnamefont {R.~F.}\ \bibnamefont {Alves}}, \bibinfo {author} {\bibfnamefont {M.}~\bibnamefont {G{\"a}rttner}}, \bibinfo {author} {\bibfnamefont {S.}~\bibnamefont {Whitlock}}, \bibinfo {author} {\bibfnamefont {G.}~\bibnamefont {Z{\"u}rn}},\ and\ \bibinfo {author} {\bibfnamefont {M.}~\bibnamefont {Weidem{\"u}ller}},\ }\bibfield  {title} {\bibinfo {title} {Glassy dynamics in a disordered {H}eisenberg quantum spin system},\ }\href@noop {} {\bibfield  {journal} {\bibinfo  {journal} {Phys. {R}ev. X}\ }\textbf {\bibinfo {volume} {11}},\ \bibinfo {pages} {011011} (\bibinfo {year} {2021})}\BibitemShut {NoStop}%
\bibitem [{\citenamefont {Schuckert}\ \emph {et~al.}(2020)\citenamefont {Schuckert}, \citenamefont {Lovas},\ and\ \citenamefont {Knap}}]{schuckert2020nonlocal}%
  \BibitemOpen
  \bibfield  {author} {\bibinfo {author} {\bibfnamefont {A.}~\bibnamefont {Schuckert}}, \bibinfo {author} {\bibfnamefont {I.}~\bibnamefont {Lovas}},\ and\ \bibinfo {author} {\bibfnamefont {M.}~\bibnamefont {Knap}},\ }\bibfield  {title} {\bibinfo {title} {Nonlocal emergent hydrodynamics in a long-range quantum spin system},\ }\href@noop {} {\bibfield  {journal} {\bibinfo  {journal} {Phys. {R}ev. {B}}\ }\textbf {\bibinfo {volume} {101}},\ \bibinfo {pages} {020416} (\bibinfo {year} {2020})}\BibitemShut {NoStop}%
\bibitem [{\citenamefont {Wootters}(1987)}]{wootters1987wigner}%
  \BibitemOpen
  \bibfield  {author} {\bibinfo {author} {\bibfnamefont {W.~K.}\ \bibnamefont {Wootters}},\ }\bibfield  {title} {\bibinfo {title} {A {W}igner-function formulation of finite-state quantum mechanics},\ }\href@noop {} {\bibfield  {journal} {\bibinfo  {journal} {Ann. {P}hys.}\ }\textbf {\bibinfo {volume} {176}},\ \bibinfo {pages} {1} (\bibinfo {year} {1987})}\BibitemShut {NoStop}%
\bibitem [{\citenamefont {Barredo}\ \emph {et~al.}(2018)\citenamefont {Barredo}, \citenamefont {Lienhard}, \citenamefont {De~Leseleuc}, \citenamefont {Lahaye},\ and\ \citenamefont {Browaeys}}]{barredo2018synthetic}%
  \BibitemOpen
  \bibfield  {author} {\bibinfo {author} {\bibfnamefont {D.}~\bibnamefont {Barredo}}, \bibinfo {author} {\bibfnamefont {V.}~\bibnamefont {Lienhard}}, \bibinfo {author} {\bibfnamefont {S.}~\bibnamefont {De~Leseleuc}}, \bibinfo {author} {\bibfnamefont {T.}~\bibnamefont {Lahaye}},\ and\ \bibinfo {author} {\bibfnamefont {A.}~\bibnamefont {Browaeys}},\ }\bibfield  {title} {\bibinfo {title} {Synthetic three-dimensional atomic structures assembled atom by atom},\ }\href@noop {} {\bibfield  {journal} {\bibinfo  {journal} {Nature}\ }\textbf {\bibinfo {volume} {561}},\ \bibinfo {pages} {79} (\bibinfo {year} {2018})}\BibitemShut {NoStop}%
\bibitem [{\citenamefont {Takei}\ \emph {et~al.}(2016)\citenamefont {Takei}, \citenamefont {Sommer}, \citenamefont {Genes}, \citenamefont {Pupillo}, \citenamefont {Goto}, \citenamefont {Koyasu}, \citenamefont {Chiba}, \citenamefont {Weidem{\"u}ller},\ and\ \citenamefont {Ohmori}}]{takei2016direct}%
  \BibitemOpen
  \bibfield  {author} {\bibinfo {author} {\bibfnamefont {N.}~\bibnamefont {Takei}}, \bibinfo {author} {\bibfnamefont {C.}~\bibnamefont {Sommer}}, \bibinfo {author} {\bibfnamefont {C.}~\bibnamefont {Genes}}, \bibinfo {author} {\bibfnamefont {G.}~\bibnamefont {Pupillo}}, \bibinfo {author} {\bibfnamefont {H.}~\bibnamefont {Goto}}, \bibinfo {author} {\bibfnamefont {K.}~\bibnamefont {Koyasu}}, \bibinfo {author} {\bibfnamefont {H.}~\bibnamefont {Chiba}}, \bibinfo {author} {\bibfnamefont {M.}~\bibnamefont {Weidem{\"u}ller}},\ and\ \bibinfo {author} {\bibfnamefont {K.}~\bibnamefont {Ohmori}},\ }\bibfield  {title} {\bibinfo {title} {Direct observation of ultrafast many-body electron dynamics in an ultracold {R}ydberg gas},\ }\href@noop {} {\bibfield  {journal} {\bibinfo  {journal} {{N}at. {C}ommun.}\ }\textbf {\bibinfo {volume} {7}},\ \bibinfo {pages} {1} (\bibinfo {year} {2016})}\BibitemShut {NoStop}%
\bibitem [{\citenamefont {Sommer}\ \emph {et~al.}(2016)\citenamefont {Sommer}, \citenamefont {Pupillo}, \citenamefont {Takei}, \citenamefont {Takeda}, \citenamefont {Tanaka}, \citenamefont {Ohmori},\ and\ \citenamefont {Genes}}]{sommer2016time}%
  \BibitemOpen
  \bibfield  {author} {\bibinfo {author} {\bibfnamefont {C.}~\bibnamefont {Sommer}}, \bibinfo {author} {\bibfnamefont {G.}~\bibnamefont {Pupillo}}, \bibinfo {author} {\bibfnamefont {N.}~\bibnamefont {Takei}}, \bibinfo {author} {\bibfnamefont {S.}~\bibnamefont {Takeda}}, \bibinfo {author} {\bibfnamefont {A.}~\bibnamefont {Tanaka}}, \bibinfo {author} {\bibfnamefont {K.}~\bibnamefont {Ohmori}},\ and\ \bibinfo {author} {\bibfnamefont {C.}~\bibnamefont {Genes}},\ }\bibfield  {title} {\bibinfo {title} {Time-domain {R}amsey interferometry with interacting {R}ydberg atoms},\ }\href@noop {} {\bibfield  {journal} {\bibinfo  {journal} {Phys. {R}ev. {A}}\ }\textbf {\bibinfo {volume} {94}},\ \bibinfo {pages} {053607} (\bibinfo {year} {2016})}\BibitemShut {NoStop}%
\bibitem [{\citenamefont {Goldschmidt}\ \emph {et~al.}(2016)\citenamefont {Goldschmidt}, \citenamefont {Boulier}, \citenamefont {Brown}, \citenamefont {Koller}, \citenamefont {Young}, \citenamefont {Gorshkov}, \citenamefont {Rolston},\ and\ \citenamefont {Porto}}]{goldschmidt2016anomalous}%
  \BibitemOpen
  \bibfield  {author} {\bibinfo {author} {\bibfnamefont {E.~A.}\ \bibnamefont {Goldschmidt}}, \bibinfo {author} {\bibfnamefont {T.}~\bibnamefont {Boulier}}, \bibinfo {author} {\bibfnamefont {R.~C.}\ \bibnamefont {Brown}}, \bibinfo {author} {\bibfnamefont {S.~B.}\ \bibnamefont {Koller}}, \bibinfo {author} {\bibfnamefont {J.~T.}\ \bibnamefont {Young}}, \bibinfo {author} {\bibfnamefont {A.~V.}\ \bibnamefont {Gorshkov}}, \bibinfo {author} {\bibfnamefont {S.}~\bibnamefont {Rolston}},\ and\ \bibinfo {author} {\bibfnamefont {J.~V.}\ \bibnamefont {Porto}},\ }\bibfield  {title} {\bibinfo {title} {Anomalous broadening in driven dissipative {R}ydberg systems},\ }\href@noop {} {\bibfield  {journal} {\bibinfo  {journal} {Phys. {R}ev. {L}ett.}\ }\textbf {\bibinfo {volume} {116}},\ \bibinfo {pages} {113001} (\bibinfo {year} {2016})}\BibitemShut {NoStop}%
\bibitem [{\citenamefont {Boulier}\ \emph {et~al.}(2017)\citenamefont {Boulier}, \citenamefont {Magnan}, \citenamefont {Bracamontes}, \citenamefont {Maslek}, \citenamefont {Goldschmidt}, \citenamefont {Young}, \citenamefont {Gorshkov}, \citenamefont {Rolston},\ and\ \citenamefont {Porto}}]{boulier2017spontaneous}%
  \BibitemOpen
  \bibfield  {author} {\bibinfo {author} {\bibfnamefont {T.}~\bibnamefont {Boulier}}, \bibinfo {author} {\bibfnamefont {E.}~\bibnamefont {Magnan}}, \bibinfo {author} {\bibfnamefont {C.}~\bibnamefont {Bracamontes}}, \bibinfo {author} {\bibfnamefont {J.}~\bibnamefont {Maslek}}, \bibinfo {author} {\bibfnamefont {E.}~\bibnamefont {Goldschmidt}}, \bibinfo {author} {\bibfnamefont {J.}~\bibnamefont {Young}}, \bibinfo {author} {\bibfnamefont {A.~V.}\ \bibnamefont {Gorshkov}}, \bibinfo {author} {\bibfnamefont {S.}~\bibnamefont {Rolston}},\ and\ \bibinfo {author} {\bibfnamefont {J.~V.}\ \bibnamefont {Porto}},\ }\bibfield  {title} {\bibinfo {title} {Spontaneous avalanche dephasing in large {R}ydberg ensembles},\ }\href@noop {} {\bibfield  {journal} {\bibinfo  {journal} {Phys. {R}ev. {A}}\ }\textbf {\bibinfo {volume} {96}},\ \bibinfo {pages} {053409} (\bibinfo {year} {2017})}\BibitemShut {NoStop}%
\bibitem [{\citenamefont {Aman}\ \emph {et~al.}(2016)\citenamefont {Aman}, \citenamefont {DeSalvo}, \citenamefont {Dunning}, \citenamefont {Killian}, \citenamefont {Yoshida},\ and\ \citenamefont {Burgd{\"o}rfer}}]{aman2016trap}%
  \BibitemOpen
  \bibfield  {author} {\bibinfo {author} {\bibfnamefont {J.}~\bibnamefont {Aman}}, \bibinfo {author} {\bibfnamefont {B.~J.}\ \bibnamefont {DeSalvo}}, \bibinfo {author} {\bibfnamefont {F.}~\bibnamefont {Dunning}}, \bibinfo {author} {\bibfnamefont {T.}~\bibnamefont {Killian}}, \bibinfo {author} {\bibfnamefont {S.}~\bibnamefont {Yoshida}},\ and\ \bibinfo {author} {\bibfnamefont {J.}~\bibnamefont {Burgd{\"o}rfer}},\ }\bibfield  {title} {\bibinfo {title} {Trap losses induced by near-resonant {R}ydberg dressing of cold atomic gases},\ }\href@noop {} {\bibfield  {journal} {\bibinfo  {journal} {Phys. {R}ev. {A}}\ }\textbf {\bibinfo {volume} {93}},\ \bibinfo {pages} {043425} (\bibinfo {year} {2016})}\BibitemShut {NoStop}%
\bibitem [{\citenamefont {Wurtz}\ \emph {et~al.}(2018)\citenamefont {Wurtz}, \citenamefont {Polkovnikov},\ and\ \citenamefont {Sels}}]{wurtz2018cluster}%
  \BibitemOpen
  \bibfield  {author} {\bibinfo {author} {\bibfnamefont {J.}~\bibnamefont {Wurtz}}, \bibinfo {author} {\bibfnamefont {A.}~\bibnamefont {Polkovnikov}},\ and\ \bibinfo {author} {\bibfnamefont {D.}~\bibnamefont {Sels}},\ }\bibfield  {title} {\bibinfo {title} {Cluster truncated {W}igner approximation in strongly interacting systems},\ }\href@noop {} {\bibfield  {journal} {\bibinfo  {journal} {Ann. {P}hys.}\ }\textbf {\bibinfo {volume} {395}},\ \bibinfo {pages} {341} (\bibinfo {year} {2018})}\BibitemShut {NoStop}%
\bibitem [{\citenamefont {Mink}\ \emph {et~al.}(2022)\citenamefont {Mink}, \citenamefont {Petrosyan},\ and\ \citenamefont {Fleischhauer}}]{mink2022hybrid}%
  \BibitemOpen
  \bibfield  {author} {\bibinfo {author} {\bibfnamefont {C.~D.}\ \bibnamefont {Mink}}, \bibinfo {author} {\bibfnamefont {D.}~\bibnamefont {Petrosyan}},\ and\ \bibinfo {author} {\bibfnamefont {M.}~\bibnamefont {Fleischhauer}},\ }\bibfield  {title} {\bibinfo {title} {Hybrid discrete-continuous truncated {W}igner approximation for driven, dissipative spin systems},\ }\href@noop {} {\bibfield  {journal} {\bibinfo  {journal} {Phys. Rev. Res.}\ }\textbf {\bibinfo {volume} {4}},\ \bibinfo {pages} {043136} (\bibinfo {year} {2022})}\BibitemShut {NoStop}%
\bibitem [{\citenamefont {Bloch}\ \emph {et~al.}(2012)\citenamefont {Bloch}, \citenamefont {Dalibard},\ and\ \citenamefont {Nascimbene}}]{bloch2012quantum}%
  \BibitemOpen
  \bibfield  {author} {\bibinfo {author} {\bibfnamefont {I.}~\bibnamefont {Bloch}}, \bibinfo {author} {\bibfnamefont {J.}~\bibnamefont {Dalibard}},\ and\ \bibinfo {author} {\bibfnamefont {S.}~\bibnamefont {Nascimbene}},\ }\bibfield  {title} {\bibinfo {title} {Quantum simulations with ultracold quantum gases},\ }\href@noop {} {\bibfield  {journal} {\bibinfo  {journal} {Nat. {P}hys,}\ }\textbf {\bibinfo {volume} {8}},\ \bibinfo {pages} {267} (\bibinfo {year} {2012})}\BibitemShut {NoStop}%
\bibitem [{\citenamefont {Orioli}\ \emph {et~al.}(2018)\citenamefont {Orioli}, \citenamefont {Signoles}, \citenamefont {Wildhagen}, \citenamefont {G{\"u}nter}, \citenamefont {Berges}, \citenamefont {Whitlock},\ and\ \citenamefont {Weidem{\"u}ller}}]{orioli2018relaxation}%
  \BibitemOpen
  \bibfield  {author} {\bibinfo {author} {\bibfnamefont {A.~P.}\ \bibnamefont {Orioli}}, \bibinfo {author} {\bibfnamefont {A.}~\bibnamefont {Signoles}}, \bibinfo {author} {\bibfnamefont {H.}~\bibnamefont {Wildhagen}}, \bibinfo {author} {\bibfnamefont {G.}~\bibnamefont {G{\"u}nter}}, \bibinfo {author} {\bibfnamefont {J.}~\bibnamefont {Berges}}, \bibinfo {author} {\bibfnamefont {S.}~\bibnamefont {Whitlock}},\ and\ \bibinfo {author} {\bibfnamefont {M.}~\bibnamefont {Weidem{\"u}ller}},\ }\bibfield  {title} {\bibinfo {title} {Relaxation of an isolated dipolar-interacting {R}ydberg quantum spin system},\ }\href@noop {} {\bibfield  {journal} {\bibinfo  {journal} {Phys. {R}ev. {L}ett.}\ }\textbf {\bibinfo {volume} {120}},\ \bibinfo {pages} {063601} (\bibinfo {year} {2018})}\BibitemShut {NoStop}%
\bibitem [{\citenamefont {Lee}\ \emph {et~al.}(2011)\citenamefont {Lee}, \citenamefont {H{\"a}ffner},\ and\ \citenamefont {Cross}}]{lee2011antiferromagnetic}%
  \BibitemOpen
  \bibfield  {author} {\bibinfo {author} {\bibfnamefont {T.~E.}\ \bibnamefont {Lee}}, \bibinfo {author} {\bibfnamefont {H.}~\bibnamefont {H{\"a}ffner}},\ and\ \bibinfo {author} {\bibfnamefont {M.}~\bibnamefont {Cross}},\ }\bibfield  {title} {\bibinfo {title} {Antiferromagnetic phase transition in a nonequilibrium lattice of {R}ydberg atoms},\ }\href@noop {} {\bibfield  {journal} {\bibinfo  {journal} {Phys. {R}ev. {A}}\ }\textbf {\bibinfo {volume} {84}},\ \bibinfo {pages} {031402} (\bibinfo {year} {2011})}\BibitemShut {NoStop}%
\bibitem [{\citenamefont {Keating}\ \emph {et~al.}(2015)\citenamefont {Keating}, \citenamefont {Cook}, \citenamefont {Hankin}, \citenamefont {Jau}, \citenamefont {Biedermann},\ and\ \citenamefont {Deutsch}}]{keating2015robust}%
  \BibitemOpen
  \bibfield  {author} {\bibinfo {author} {\bibfnamefont {T.}~\bibnamefont {Keating}}, \bibinfo {author} {\bibfnamefont {R.~L.}\ \bibnamefont {Cook}}, \bibinfo {author} {\bibfnamefont {A.~M.}\ \bibnamefont {Hankin}}, \bibinfo {author} {\bibfnamefont {Y.-Y.}\ \bibnamefont {Jau}}, \bibinfo {author} {\bibfnamefont {G.~W.}\ \bibnamefont {Biedermann}},\ and\ \bibinfo {author} {\bibfnamefont {I.~H.}\ \bibnamefont {Deutsch}},\ }\bibfield  {title} {\bibinfo {title} {Robust quantum logic in neutral atoms via adiabatic {R}ydberg dressing},\ }\href@noop {} {\bibfield  {journal} {\bibinfo  {journal} {Phys. {R}ev. {A}}\ }\textbf {\bibinfo {volume} {91}},\ \bibinfo {pages} {012337} (\bibinfo {year} {2015})}\BibitemShut {NoStop}%
\bibitem [{\citenamefont {Buchmann}\ \emph {et~al.}(2017)\citenamefont {Buchmann}, \citenamefont {M{\o}lmer},\ and\ \citenamefont {Petrosyan}}]{buchmann2017creation}%
  \BibitemOpen
  \bibfield  {author} {\bibinfo {author} {\bibfnamefont {L.}~\bibnamefont {Buchmann}}, \bibinfo {author} {\bibfnamefont {K.}~\bibnamefont {M{\o}lmer}},\ and\ \bibinfo {author} {\bibfnamefont {D.}~\bibnamefont {Petrosyan}},\ }\bibfield  {title} {\bibinfo {title} {Creation and transfer of nonclassical states of motion using {R}ydberg dressing of atoms in a lattice},\ }\href@noop {} {\bibfield  {journal} {\bibinfo  {journal} {Phys. {R}ev. {A}}\ }\textbf {\bibinfo {volume} {95}},\ \bibinfo {pages} {013403} (\bibinfo {year} {2017})}\BibitemShut {NoStop}%
\bibitem [{\citenamefont {Mitra}\ \emph {et~al.}(2020)\citenamefont {Mitra}, \citenamefont {Martin}, \citenamefont {Biedermann}, \citenamefont {Marino}, \citenamefont {Poggi},\ and\ \citenamefont {Deutsch}}]{mitra2020robust}%
  \BibitemOpen
  \bibfield  {author} {\bibinfo {author} {\bibfnamefont {A.}~\bibnamefont {Mitra}}, \bibinfo {author} {\bibfnamefont {M.~J.}\ \bibnamefont {Martin}}, \bibinfo {author} {\bibfnamefont {G.~W.}\ \bibnamefont {Biedermann}}, \bibinfo {author} {\bibfnamefont {A.~M.}\ \bibnamefont {Marino}}, \bibinfo {author} {\bibfnamefont {P.~M.}\ \bibnamefont {Poggi}},\ and\ \bibinfo {author} {\bibfnamefont {I.~H.}\ \bibnamefont {Deutsch}},\ }\bibfield  {title} {\bibinfo {title} {Robust {M}{\o}lmer-{S}{\o}rensen gate for neutral atoms using rapid adiabatic {R}ydberg dressing},\ }\href@noop {} {\bibfield  {journal} {\bibinfo  {journal} {Phys. {R}ev. A}\ }\textbf {\bibinfo {volume} {101}},\ \bibinfo {pages} {030301} (\bibinfo {year} {2020})}\BibitemShut {NoStop}%
\bibitem [{\citenamefont {Genkin}\ \emph {et~al.}(2014)\citenamefont {Genkin}, \citenamefont {W{\"u}ster}, \citenamefont {M{\"o}bius}, \citenamefont {Eisfeld},\ and\ \citenamefont {Rost}}]{genkin2014dipole}%
  \BibitemOpen
  \bibfield  {author} {\bibinfo {author} {\bibfnamefont {M.}~\bibnamefont {Genkin}}, \bibinfo {author} {\bibfnamefont {S.}~\bibnamefont {W{\"u}ster}}, \bibinfo {author} {\bibfnamefont {S.}~\bibnamefont {M{\"o}bius}}, \bibinfo {author} {\bibfnamefont {A.}~\bibnamefont {Eisfeld}},\ and\ \bibinfo {author} {\bibfnamefont {J.}~\bibnamefont {Rost}},\ }\bibfield  {title} {\bibinfo {title} {Dipole-dipole induced global motion of {R}ydberg-dressed atom clouds},\ }\href@noop {} {\bibfield  {journal} {\bibinfo  {journal} {J. {P}hys. {B}-{A}t. {M}ol. {O}pt.}\ }\textbf {\bibinfo {volume} {47}},\ \bibinfo {pages} {095003} (\bibinfo {year} {2014})}\BibitemShut {NoStop}%
\end{thebibliography}%

\appendix
\section{Classical Equation of Motion}
\label{appendix:eom}
According to the Hamiltonian~\eqref{hamiltonian}, the classical equations of motion are 
\begin{align}
\dot{r}_i^\alpha&=\frac{p_i^\alpha}{m} \\
\dot{p}_i^\alpha&=-\sum_{ j\in \text{n.n. } i}\frac{3\hbar Ja_l^6}{2r_{ij}^8}(r_i^\alpha-r_j^\alpha)(\sigma_i^z+1)(\sigma_j^z+1)  \\
&\hspace{0.15in}{}-\frac{\hbar V_0\pi^2}{a_l^2}\tilde{r}_i^\alpha(1-\sigma_i^z) \nonumber\\
\dot{\sigma_i^\alpha}&=\left[\frac{ \Delta}{2}+\!\!\!\sum_{j\in \text{n.n. } i}\frac{ Ja_l^6}{2r_{ij}^6}(\sigma_j^z+1)+\frac{ V_0\pi^2}{2a_l^2}\tilde{r}_i^2\right]\sum_\beta \epsilon^{\alpha\beta z}\sigma_i^\beta \nonumber\\
&{}-\frac{\Omega}{2}\sum_\beta\epsilon^{\alpha\beta x}\sigma_i^\beta -\frac{1}{T_1}(-1+\sigma_i^z)\delta^{\alpha z} -\frac{1}{T_2}\sigma_i^\alpha(1-\delta^{\alpha z}), \label{eq:classical-eom}
\end{align}
where $\epsilon^{ijk}$ is the Levi-Civita symbol, $\delta^{\alpha \beta}$ is the Kronecker delta, and the variables $r_i^\alpha$, $p_i^\alpha$, and $\sigma_j^z$ are  c-numbers corresponding to their quantum operator counterparts. 
In the calculation of the ramping dynamics in Sec.~\ref{sec:ramping}, we include single particle decay time $T_{1}$ and the dephasing time $T_2$ in the model and compare the effects of single-body noises with those of the atom motion. To directly compare the motional dynamics to the Ising dynamics, both models exclude factors that may change $T_{1}$ and $T_{2}$, such as Doppler shift, AC Stark shift, magnetic field inhomogeneity and interplay between motion and single-particle noises. We expect these factors to be less  impactful than atom motion. We also want to point out that a more precise way to incorporating the dissipative terms in TWA methods is recently developed in \cite{mink2022hybrid}.

\section{Dependence of correlations, kinetic energy, and atom number on time}
\label{appendix:time}
\begin{figure}[h]
    \centering
    \includegraphics[width=0.85\columnwidth]{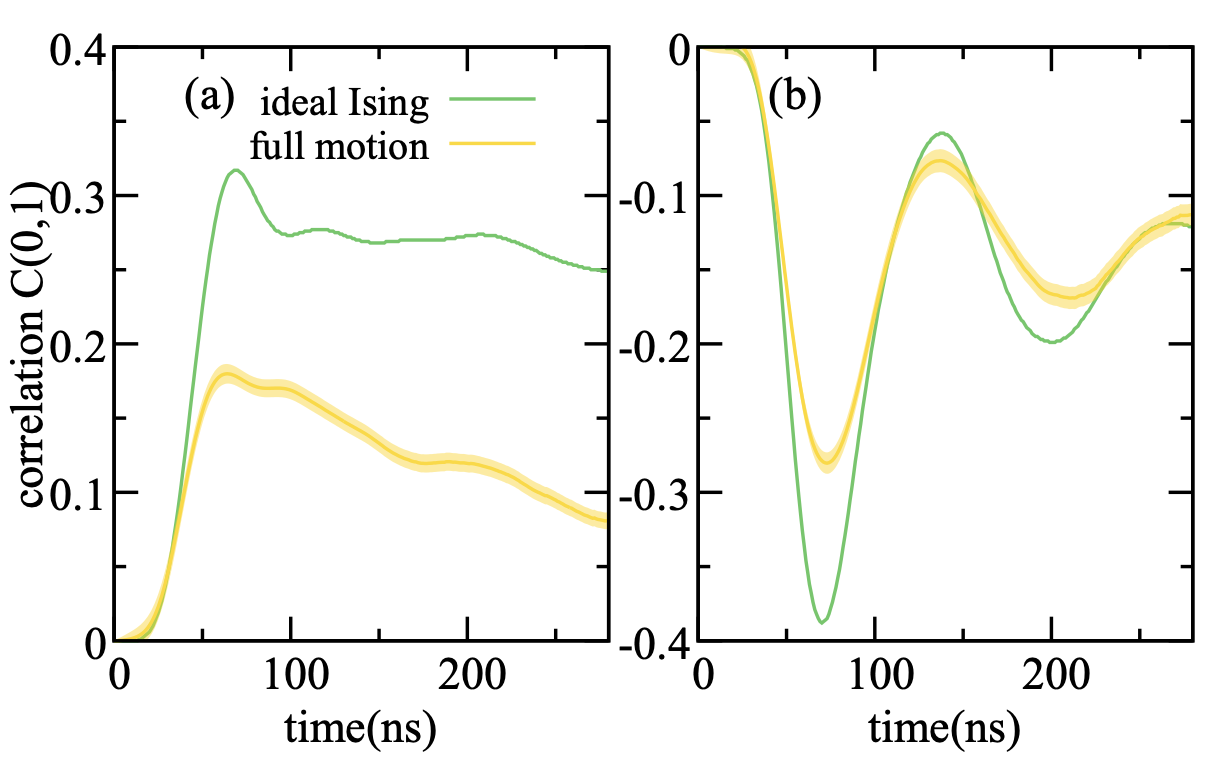}
    \caption{Nearest-neighbor correlation C(0,1) changes over time (a)  off-resonance ($\Delta=-10$MHz) and (b) at the single-atom resonance ($\Delta=0$MHz). }
    \label{fig:time}
\end{figure}

To see how the effect of motion evolves, in Fig.~\ref{fig:time} we show the values of the nearest-neighbor correlation $C(0,1)$ over time in the long-time quench at $\Delta=-10$MHz (far from the single-particle resonance)  and $\Delta=0$MHz (on resonance). These roughly are the  detunings where atom motion has the largest and smallest effects, respectively. 
Fig.~\ref{fig:time} shows that the effect of motion accumulates over time, as the ideal Ising and full motion  curves  overlap at the beginning of the quench, but start to show differences after $t=50$ns.

\begin{figure}
    \includegraphics[width=0.5\columnwidth]{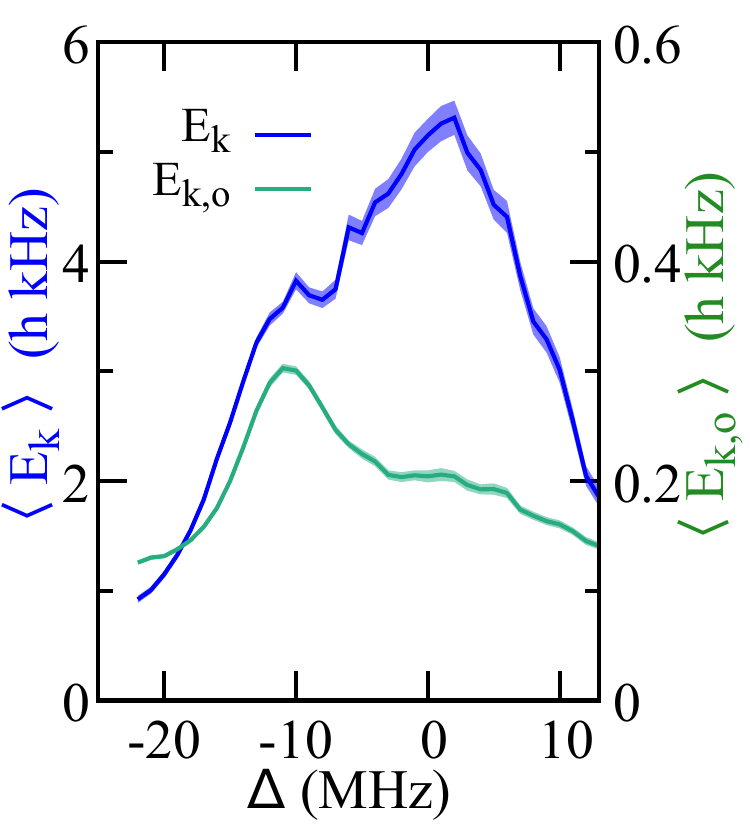}
    \caption{\label{fig:mot_effect} Average kinetic energy (blue curve) and out-of-plane kinetic energy (green curve) at the end of the long-time quench  as a function $\Delta$. 
    Initially, the zero-point energy per direction is $\hbar\omega/2=h\times 0.108$kHz. 
    }
\end{figure}

To  gain  insight  into  the  role  that  atom motion  plays in the   dynamics,   we examine the motional behavior directly. The average kinetic energy in Fig.~\ref{fig:mot_effect} suggests that, when allowed to move, the kinetic energy of atoms will  increase, especially in the in-plane direction. This is consistent with the effects of motional decoherence increasing over time during the quench. 

\begin{figure}
    \includegraphics[width=0.5\columnwidth]{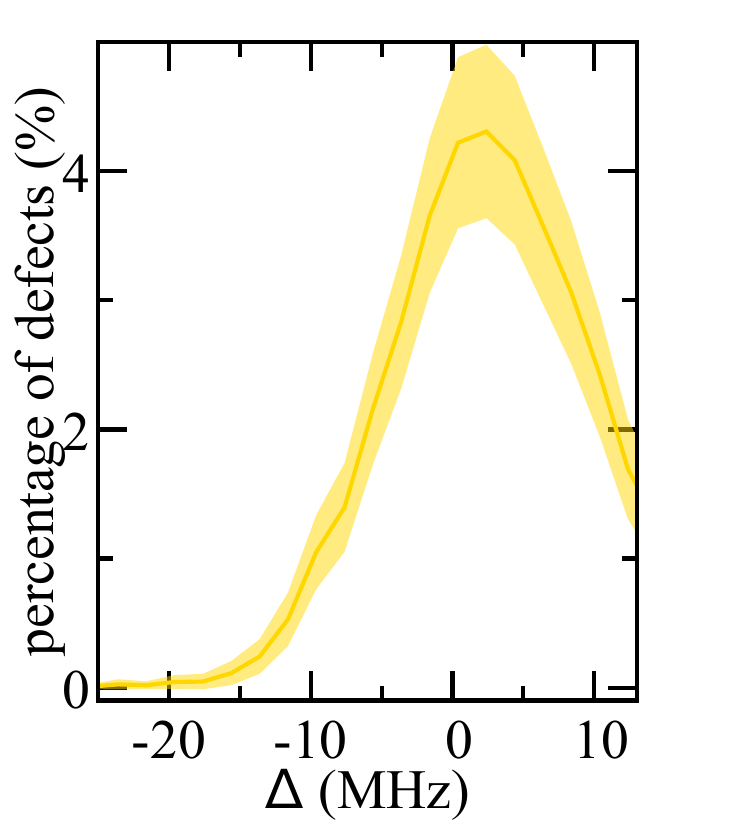}
    \caption{\label{fig:defect} Percentage of atom loss at the end of the long-time quench as a function of  $\Delta$.}
\end{figure}

Also, at long enough time some Rydberg atoms will  get close enough to each other and react, leading to atom losses. During the dynamics, some atoms can move out of the traps and cause atom loss as well. At the end of dynamics, all lost atoms are treated as Rydberg-state atoms, according to the experimental detection techniques in Ref.~\cite{guardado2018probing}.  However, as plotted in Fig.~\ref{fig:defect}, after the long-time quench in Sec.~\ref{sec:long}, the fraction of motion-induced atoms loss by the end of the dynamic is $<5\%$,   a small impact on the simulation results shown.

\nocite{*}


\end{document}